\documentclass{emulateapj}

\usepackage{amsmath}
\usepackage{graphicx}
\usepackage{float}
\usepackage{array, booktabs}
\usepackage{natbib}

\newcommand{\Eqref}[1]{Eq.\eqref{#1}}
\newcommand{\Eqsref}[1]{Eqs.\eqref{#1}}
\newcommand{\figref}[1]{Figure~\ref{#1}}
\newcommand{\figsref}[1]{Figures~\ref{#1}}
\newcommand{\tabref}[1]{Tab.\ref{#1}}
\newcommand{\Ni}{$^{56}{\rm Ni}$~}

\begin{document}
	
			\title{A collapsar model with  disk wind: Implications for supernovae associated with Gamma-Ray Bursts}

			\author{Tomoyasu Hayakawa}
			\affiliation{Department of astronomy, Kyoto University, Kitashirakawa-Oiwake-cho, Sakyo-ku, Kyoto 606-8502, Japan}
			\email{hayakawa@kusastro.kyoto-u.ac.jp}

			\author{Keiichi Maeda}
			\affiliation{Department of astronomy, Kyoto University, Kitashirakawa-Oiwake-cho, Sakyo-ku, Kyoto 606-8502, Japan}
			\email{keiichi.maeda@kusastro.kyoto-u.ac.jp}

			\received{22th, Nov, 2017}

			\begin{abstract}
	　We construct a simple but self-consistent collapsar model for gamma-ray bursts (GRBs) and SNe associated with GRBs (GRB-SNe). 
	Our model includes a black hole, an accretion disk, and the envelope surrounding the central system.
	The evolutions of the different components are connected by the transfer of the mass and angular momentum. 
	To address properties of the jet and the wind-driven SNe, we consider competition of the ram pressure from the infalling envelope and those from the jet and wind. 
	The expected properties of the GRB jet and the wind-driven SN are investigated as a function of the progenitor mass and angular momentum. 
	We find two conditions which should be satisfied if the wind-driven explosion is to explain the properties of the observed GRB-SNe. (1) The wind should be collimated at its base, and (2) it should not prevent further accretion even after the launch of the SN explosion. Under these conditions, some relations seen in the properties of the GRB-SNe could be reproduced by a sequence of different angular momentum in the progenitors.
	Only the model with the largest angular momentum could explain the observed (energetic) GRB-SNe, and we expect that the collapsar model can result in a wide variety of observational counterparts mainly depending on the angular momentum of the progenitor star. 

			\end{abstract}

			\section{Introduction}

	A gamma-ray burst (GRB) is one of the most energetic events in the universe.
	GRBs release gamma-rays typically in a few seconds, with the total isotropic radiation energy reaching $\sim 10^{53}-10^{54}{\rm erg}$.
	GRBs show diversities in their total energies, durations, and their spectra \citep{Piran_2004,Meszaros_2006,Woosley_Bloom_2006}.
	In this paper, we focus on the population of GRBs whose durations exceed $\sim$ 2 seconds, which are called long-GRBs \citep{Kouveliotou_et_al_1993}.
	They are believed to originate from a core-collapse event of massive stars at the end of their lives.

	The collapsar model is a leading scenario for long-GRBs, in which the central engine is a system of a black hole and an accretion disk formed through the collapse of a fast rotating massive star \citep{Woosley1993,MacFadyenWoosley1999,Pophametal1999}.
	The accretion disk around the black hole is hyper accreting, then the gravitational energy would be extracted to launch a relativistic jet.
	If the jet could penetrate all the way through the stellar envelope, it would produce high energy radiations observed as a long GRB.

	Long GRBs are sometimes accompanied by supernovae (SNe).
	An SN associated with a GRB (here after GRB-SN) was first detected in 1998: SN1998bw within the error box of GRB980425 \citep{Galamaetal1998}.
	This SN is of type Ic (SNe Ic) 
		\footnote[1] 
		{SNe Ic are classified by no hydrogen and  helium absorption lines in their spectra \citep{Filippenko_1997}. 
		It has been interpreted that they originate from a core collapse of a massive star that had lost their hydrogen and helium envelopes before the explosion \citep{Nomoto_etal_1994,Woosley_Langer_Weaver_1995}.}
	and its kinetic energy is estimated to be larger than typical SNe by more than an order of magnitude \citep{Iwamotoetal1998,Woosley_etal_1999,Maeda_etal_2006}.
	There are an increasing number of detections of such hyper energetic SNe associated with GRBs \citep[see reviews][and references therein]{Wang_Wheeler_1998,Nomoto_etal_2006,Bissaldi_etal_2007,valle_2011,Cano_2013,Hjorth_2013}.
	These SNe with a large kinetic energy are sometimes called `hypernovae', or observationally `broad-lined SNe Ic (SNe Ic-BL)' based on their spectra showing characteristics of SNe Ic but with broad absorption lines.

	Any models for the central engine should be able to explain the GRB-SN association, and further relations between the natures of GRBs and associated SNe.
	Addressing this association requires a self-consistent treatment of the central engine and the envelope of the collapsing stellar envelope, since (1) the GRB central engine is powered by accretion of the envelope, and (2) the SN is  regarded to be successful only when a large fraction of the collapsing envelope are ejected.

	The original collapsar model calculation in 2D was performed by \cite{MacFadyenWoosley1999}.
	They focused on the core of the massive star, not including most of the envelope in their simulation grids.
	They suggested that the jet found in their simulation is able to explain the main characteristics of observed GRBs.
	As for the associated SN, they suggested that a disk wind launched from the vicinity of the hyper-accreting BH could be a source of the SN.
	However, interaction between the disk wind and stellar envelope has not been investigated, which limits a proper comparison between the model and the observed features of GRB-SNe.

	\cite{Kohrietal2005} concluded that the kinetic energy of the wind from the hyper-accreting disk could reach to a few $\times 10^{52}{\rm erg}$.
	However, they considered only the black hole and the disk, without considering the action of the stellar envelope.

	\cite{Kumaretal2008} were the first who constructed a simple collapsar model including the black hole, the disk and the stellar envelope.
	Their motivation was to explain the natures of GRBs and their X-ray afterglows based on a Wolf-Rayet progenitor in the collapsar model.
	As a by-product they found that the wind energy can reach to a few $10^{52}{\rm erg}$, assuming that this disk wind can immediately blow the envelope following the launch.
	They also discussed how the jet energy is affected by the angular momentum of the progenitor, but a relation between the disk wind and the nature of the progenitor was not discussed.
	Therefore, a possible relation between the natures of the GRB jet and the wind, and how it is connected by the nature of the progenitor, have not been clarified.

	Our motivation in this paper is to find expected relations in the natures of GRBs and GRB-SNe by a simple model under the collapsar scenario including the black hole, the disk and the stellar envelope.
	Especially, we explore whether the disk wind could blow the stellar envelope to produce observed GRB-SNe, by taking into account the action of the collapsing stellar envelope onto the disk wind.
	In addition, by using different progenitors, we aim to explore relations between the nature of the progenitors, GRBs and GRB-SNe.

	The paper is structured as follows.
	In \S 2, we introduce the method of our simulation.
	Our results are then described for different configurations for the disk wind: a spherical wind (\S 3) and a collimated wind (\S 4). 
	Further parameter surveys on the nature of the progenitor star are presented in \S 5.
	In \S 6, implications are discussed on the nature of expected observational counterparts.
	In \S 7, we summarize our findings and close the paper.

	\section{Method}

\begin{figure}
				\includegraphics*[width = \columnwidth, bb = 0 0 1024 768]{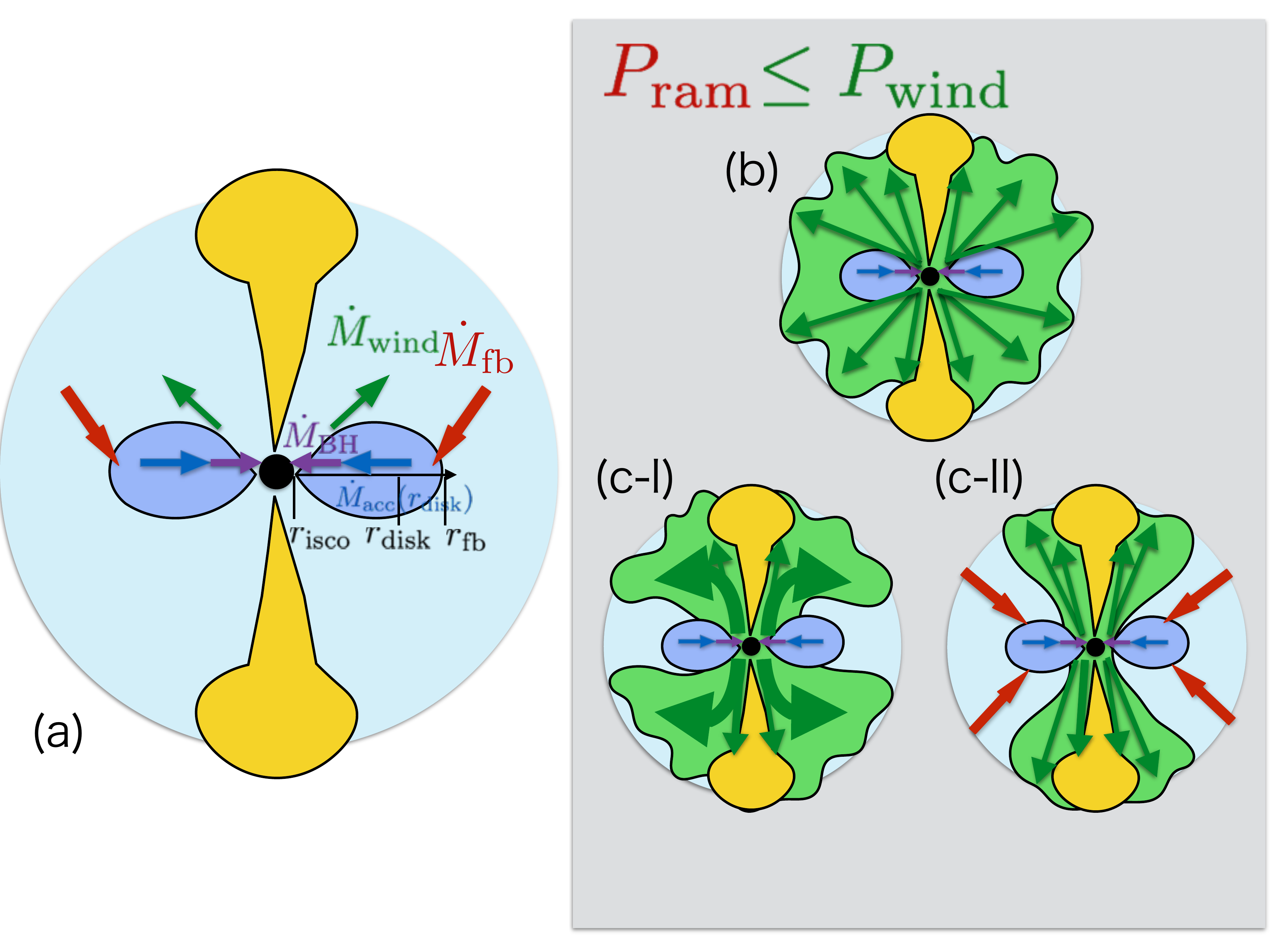}
				\caption{Cartoons which illustrate our simulation setups. The BH, disk, envelope are shown by black, blue, and cyan, respectively. 
				The GRB jet is shown by the yellow region, while the disk wind is shown by the green region. 
				The situations before and after the launch of the wind-driven SN are shown in (a) and (b, c-I, c-II), respectively. 
				The treatment of the wind is tested for three situations, (b) a quasi-spherical wind, (c-I) a collimated but laterally expanding wind (Case OUT), and (c-II) a collimated wind with continuous accretion to the central system (Case IN-OUT). See the text for details.}
				 \label{cartoon_1}
			\end{figure}

	In this section, we describe our method for this survey.
	We construct a simple but self-consistent collapsar model by taking into account all of the black hole, the disk and the stellar envelope which are mutually connected by transfer of mass and angular momentum (\figref{cartoon_1}).
	Our method largely follows prescriptions given by \cite{Kumaretal2008}.

	In addition, we take into account a disk wind in a parameterized way, which allows us to examine the condition for the launch of an SN through  competition  between the ram pressure of the wind and that of the envelope.
	If the disk wind could overwhelm  the ram pressure of the infalling material, we regard that an SN explosion is launched.
	The jet energy is also estimated, by taking into account  the pressure competition.
	
	The pre-collapse structure is taken from \cite{WoosleyHeger2006}.
	This progenitor model sequence includes mostly rapidly rotating Wolf-Rayet stars.
	While we use a specific model 16TI, we change the total angular momentum and disk wind parameter $s_{\rm g}$ in our parameter survey (see $\S$ \ref{cal_disk_wind} for more details).

	\subsection{Prior to the disk formation}

		The core collapse is initiated when pressure support is lost after the formation of the iron core.
		We assume there is no SN explosion at the formation of a neutron star (NS), and the NS continues to accrete the infalling matter and becomes a black hole (BH).
		At the beginning of the calculation, the central region is replaced by a point mass representing the newly formed BH.
		Initially, the infalling matter in the adopted pre-collapse models does not have enough angular momentum to form a disk around the BH.
		In this initial phase, the BH evolution is approximately described by free fall.

		We divide the progenitor into $N$ shells ($N = 708$), where each shell has its angular momentum and mass according to the initial condition of the progenitor star.
		We define physical quantities in the $i$-th shell by the subscript $i$.
		The free fall time of the $i$-th shell, denoted by $t_{{\rm ff}, i}$, is given as follows.	
			\begin{align}
				t_{{\rm ff}, i} &\simeq  \frac{\pi}{2\sqrt{2}}\frac{1}{\Omega_{{\rm K},i}(r_{0 ,i})} \left(1 + \frac{3}{8}\left(\frac{\Omega_{0,i}(r_{0,i})}{\Omega_{{\rm K},i}(r_{0,i})}\right)^{2}\right). \label{t_ff}
			\end{align}
		Here, the subscript $0$ refers to the initial condition.
		$\Omega_{{\rm K},i}(r_{0,i})$ is the Kepler rotation angular velocity, $\Omega_{0,i}(r_{0, i})$ is the angular velocity of the $i$-th shell in the initial pre-collapse condition, and $r_{0,i}$ is the initial location of the $i$-th shell from the center of the progenitor.
		To connect $t_{{\rm ff}, i}$ to the time through which the $i$-th shell collapses to the BH ($t_{{\rm eq}, i}$), the sound crossing time ($t_{{\rm sound}, i}$) must be added.
		In the observer frame, these time scales are given as follows;
			\begin{align}
				t_{{\rm sound},i} &\simeq \sum_{k=1}^{i}\frac{r_{0 ,k} -r_{0 , k-1}}{\sqrt{P_{0,k}/\rho_{0,k}}}, \\
				t_{{\rm eq} ,i} &= t_{{\rm ff},i} + t_{{\rm sound},i}.
			\end{align}
		$P_{0,k}$ and $\rho_{0,k}$ are the initial (pre-collapse) pressure and density, respectively, in the $k$-th shell.
		Between $ t_{{\rm eq},i-1} \leq t \leq t_{{\rm eq},i}$, the accretion rate onto the BH, $\dot{M}_{\rm BH}$, is approximately estimated as following;
			\begin{align}
				\dot{M}_{\rm fb}(t) &= \frac{m_{{\rm shell},i}}{\delta t_{\rm eq}} ~~ , \label{dot_m_fall_back}\\
				\delta t_{{\rm eq} ,i} &= t_{{\rm eq} ,i} - t_{{\rm eq} , i-1} ~~ (t_{{\rm eq},i-1}\leq t \leq t_{{\rm eq},i}).
			\end{align}
		Here, $\dot{M}_{\rm fb}(t)$ is an accretion rate from the envelope to the central system consisting of a BH and a disk (if present).
		In the initial phase,
			\begin{align}
				\dot{M}_{\rm BH}(t) &= \dot{M}_{\rm fb}(t).
				\label{bh_mass_evo_before_disk}
			\end{align}
		$\dot{M}_{\rm fb}(t)$ changes as a function of time, which is determined only by the initial condition of the progenitor.
		
		The above \Eqref{bh_mass_evo_before_disk} is only applicable when the envelope directly collapses to the BH.
		Given the increasing specific angular momentum outward in the progenitor star model, at some point the centrifugal force could be sufficiently strong to stop the direct collapse, leading to the disk formation.
		To judge when the transition happens, the key quantity is the Keplerian radius ($r_{{\rm fb}, i}$) for each shell; 
			\begin{align}
				r_ {{\rm fb},i}(t) = r_{0,i}\left( \frac{\Omega_{0,i}(r_{0.i})}{\Omega_{{\rm K},i}(r_{0,i})}\right)^{2} (t_{{\rm eq},i-1} \leq t \leq t_{{\rm eq}, i}).
				\label{r_fb}
			\end{align}
		Approximately, the $i$-th shell falls toward this radius ($r_{\rm fb}$) in the time duration $\delta t_{{\rm eq}, i}$.
		We estimate that the disk formation takes place after the infalling Keplerian radius becomes larger than the innermost stable circular orbit (ISCO) radius of the central BH ($r_{\rm isco}(t)$), i.e., $r_{{\rm fb}, i}(t) > r_{\rm isco}(t)$.
		
		We can estimate the angular momentum evolution of the BH in the initial phase, as follows;
			\begin{align}
				 \dot{J}_{\rm fb}(t) &= \frac{J_{{\rm shell}, i}}{\delta t_{{\rm eq}, i}} ~~\left( t_{{\rm eq}, i-1} \leq t \leq t_{{\rm eq}}\right),\label{dot_J_fb}\\
				J_{\rm BH}(t) &= \int_{0}^{t} \dot{J}_{\rm fb}(t^{\prime})dt^{\prime} = \sum_{k=1}^{i} J_{{\rm shell}, k} ~~\left( t_{{\rm eq}, i-1} \leq t \leq t_{{\rm eq}, i}\right).
				\label{J_pre_disk}
			\end{align}
		$J_{{\rm shell}, i}$ is the angular momentum of the $i$-th shell, and $\dot{J}_{\rm fb}(t)$ is the transfer rate of the angular momentum from the $i$-th shell in the envelope to the central system.

	\subsection{After the disk formation}
		If $r_{{\rm fb}, i}(t) \geq r_{\rm isco}(t)$, the collapsing gases from the envelope begin to fall onto a disk. 
		The accretion rate and angular momentum transport onto the disk from the envelope can be  estimated by the same method described in $\S2.1$ (e.g. \Eqsref{dot_m_fall_back} and \eqref{dot_J_fb}).
		
		Once the disk is formed, accretion time scale is dominated by viscous evolution of the disk.
		The disk is supposed to keep nearly Kepler rotation with viscosity dissipation.
		We thus use Shakura Sunyaev's ~$\alpha$-model \citep{ShakuraSunyaev1973}.
		The typical size of the disk ($r_{\rm disk}(t)$) is estimated as follows;
			\begin{align}
				r_{\rm disk}(t) &= \frac{\left(\frac{J_{\rm disk}(t)}{M_{\rm disk}(t)}\right)^{2}}{GM_{\rm BH}(t)}.
				\label{r_disk}
			\end{align}
		Then the viscous time scale ($t_{\rm acc}(t)$) and the disk accretion rate onto the BH at $r_{\rm disk}(t)$ are estimated as follows;
			\begin{align}
				t_{\rm acc}(t) &\simeq \frac{2}{\alpha\sqrt{\frac{GM_{\rm BH}(t)}{r_{\rm disk}^3(t)}}},\label{t_acc}\\
				\dot{M}_{\rm acc}(t, r_{\rm disk}) &\simeq \frac{M_{\rm disk}(t)}{t_{\rm acc}(t)},\label{dotM_acc}
			\end{align}
		where $\alpha$ is the viscosity parameter (for which we adopt $\alpha = 0.1$), $\dot{M}_{\rm acc}(t, r_{\rm disk})$ is the accretion rate in the disk.
		$M_{\rm disk}(t)$ and $J_{\rm disk}(t)$ are the disk total mass and angular momentum, respectively. 
		Here, $\dot{M}_{\rm acc}(t, r_{\rm disk})$ is not necessarily equal to the accretion rate onto the BH, because the disk wind could take a part of the mass away from the disk while the remaining fraction accretes onto the BH.
		The treatment for this effect is discussed in the next section.
		We introduce the net accretion rate by using the mass loss rate of the wind as follow;
			\begin{align}
				\dot{M}_{\rm BH}(t) &= \dot{M}_{\rm acc}(t, r_{\rm disk}) - \dot{M}_{\rm wind}(t)\label{dot_m_bh_kari}.
			\end{align}
		
		In the first stage of the disk evolution, $t_{\rm acc}(t)$ becomes very small because $r_{\rm disk}(t)$ is small.
		In this situation, we regard that the accretion rate from the disk to the BH is equal to that from the envelope to the disk.
		If $M_{\rm disk}(t) \simeq 0$ and $J_{\rm disk}(t) \simeq 0$, then
			\begin{align}
				\frac{J_{\rm disk}(t)}{M_{\rm disk}(t)} &\simeq \frac{J_{{\rm shell},i}}{M_{{\rm shell},i}}, ~{\rm and}~ r_{\rm disk}(t) \simeq r_{{\rm fb},i},\
			\end{align}
		and if $t_{\rm acc} \ll \delta t_{{\rm eq},i}$ then
			\begin{align}
				\dot{M}_{\rm acc}(t, r_{\rm disk}) &\simeq \dot{M}_{\rm fb}(t).
			\end{align}
		The rate of the angular moment transport can be estimated in a similar way, by considering the BH spin and general relativity.
		To calculate this one, we follow \cite{Bardeen_etal_1972}.
		Denoting the BH spin parameter as $a$ and the Shwarzschild radius as $r_{\rm s}$, we use the following expressions;

			\begin{align}
					a(t) &= \frac{cJ_{\rm BH}(t)}{GM_{\rm BH}^{2}(t)} , ~ r_{\rm s}(t) = \frac{2GM_{\rm BH}(t)}{c^{2}}\\
					z_{1}(t) &= 1 + \left(1-a^{2}(t)\right)^{\frac{1}{3}}\left( \left(1+a(t)\right)^{\frac{1}{3}} + \left( 1-a(t)\right)^{\frac{1}{3}}\right),\\
					z_{2}(t) &= \left( 3a^{2} + z_{1}^{2}\right)^{\frac{1}{2}},\\
					r_{\rm isco}(t) &= \frac{GM_{\rm BH}(t)}{c^{2}}\left(3 + z_{2} - \sqrt{\left(3-z_{1} \right)\left( 3+z_{1} +2z_{2}\right)} ~\right),\\
					j_{\rm isco}(t) &= \sqrt{GM_{\rm BH}(t) r_{\rm isco}(t)}\times \nonumber\\
					&\frac{ r_{\rm isco}^{2} -a r_{\rm s}\sqrt{\frac{r_{\rm isco}r_{\rm s}}{2}} +\left( \frac{ar_{\rm s}}{2}\right)^{2}}{ r_{\rm isco} \left( r_{\rm isco}^{2} -\frac{3}{2}r_{\rm isco}r_{\rm s} +ar_{\rm s}\sqrt{\frac{r_{\rm isco}r_{\rm s}}{2}}\right)^{\frac{1}{2}}},\\
					\dot{J}_{\rm BH}(t) &\simeq j_{\rm isco}(t) \times \dot{M}_{\rm BH}(t),
			\end{align}
		where $j_{\rm isco}$ is the specific angular momentum at the innermost stable circular orbit.
		As time goes by, the disk mass can increase because of increasing  angular momentum of the collapsing shell.
		In this situation, the disk mass evolution is estimated as follows;
			\begin{align}
				\dot{M}_{\rm disk}(t) = \dot{M}_{\rm fb}(t) - \dot{M}_{\rm acc}(t, r_{\rm disk}).\label{mass_disk_evo}
			\end{align}

		The equation describing the angular momentum is given in a similar manner.
		However, these equations require further modification (additional terms) if we are to include the jet and the disk wind.
		This is described in the subsequent sections.
		For the disk wind, we assume a part of the accretion escapes from the disk (see the next section).
		Even if we consider the mass evolution of the disk with the disk wind, \Eqref{mass_disk_evo} is not changed because the disk wind escapes from the disk as a part of the accretion material, thus the effect is already included (i.e., \Eqref{dot_m_bh_kari}).

		\subsubsection{Disk wind}\label{cal_disk_wind}
			Not all the gases given by $\dot{M}_{\rm acc}(t,r_{\rm disk})$ may accrete onto the BH, depending on the condition within the disk.
			We consider two phases as the disk state; the neutrino dominated accretion flow (NDAF) \citep{NarayanYi1994,Pophametal1999}, and the advection dominated accretion flow (here after ADAF)\citep{NarayanYi1994}.

			NDAF is an accretion flow efficiently cooled by neutrinos, which would not trigger a strong disk wind.
			On the other hand, ADAF is radiatively inefficient, and the dissipated energies are trapped into the accretion flow.
			A large fraction of the gas in the ADAF disk likely escapes from the disk as a wind rather than accreting onto the BH \citep{MacFadyenWoosley1999,Narayanetal2001,Kohrietal2005}.

			How the wind is blown from the disk is still unknown.
			We adopt a simple prescription on the wind mass loss rate as a function of radius within the disk ($\dot{M}_{\rm wind}(t,r)$), as follows \citep{Blandford_Begelman_1999,Piran_2004,Kohrietal2005};
				\begin{align}
					\dot{M}_{\rm wind}(t, r) &\simeq \dot{M}_{\rm acc}(t,r_{\rm outer})\times \left[ 1 -\left(\frac{r}{r_{\rm outer}(t)} \right)^{s}\right], ~(0 \leq s \leq 1) \label{dot_m_wind}
				\end{align}
			where $s$ is a wind parameter as described below. 
			This is the formula used by \cite{Kohrietal2005} and \cite{Yuanetal2005} in discussing the disk wind properties.
			\cite{Kohrietal2005} defined an advection factor $f_{\rm adv}(t,r)$, as follow;
				\begin{align}
					f_{\rm adv}(t,r) &\equiv \frac{q_{\rm adv}}{q_{\rm vis}} \equiv \frac{q_{\rm vis} - q_{\rm cool}}{q_{\rm vis}}, \label{f_adv}
				\end{align}
			$q_{\rm adv}, q_{\rm vis}, q_{\rm cool}$ are a rate of the energy carried by advection, a heating rate by the viscosity, and cooling rate by radiation.
			\cite{Yuanetal2005} related this advection parameter to the disk wind mass loss rate (e.g., \Eqref{dot_m_wind}) as follow;
				\begin{align}
					\frac{d\log{\dot{M}_{\rm acc}(t,r)}}{d\log{r}} &= s(t, r),\\
					\frac{d\dot{M}_{\rm acc}(t, r)}{dr} &= s(t, r)\frac{\dot{M}_{\rm acc}(t, r)}{r},\label{d_dot_m_wind}\\
					s(t, r) &= s_{0}f_{\rm adv}(t, r),\label{s_0}
				\end{align}
			where $s_{0}$ is a parameter ($0\leq s_{0} \leq 1$). 
			The wind parameter $s(t, r)$ is highly dependent on whether the disk state is NDAF or ADAF, thus implicitly on the accretion rate.
			For computing the wind properties, we assume $s_{\rm 0}$ is constant.
			Referring to Figure. 3 of \cite{Kohrietal2005}, we adopt $f_{\rm adv} = 1, ~{\rm and} ~0.4$ in the ADAF and NDAF regions, respectively.
			This prescription leads to a more efficient wind generation than \cite{Kumaretal2008}.

			In our approach, we do not solve radial distribution within the disk.
			So, we integrate  \Eqref{d_dot_m_wind} from $r_{\rm isco}$ to $r_{\rm disk}$ in order to obtain the total mass outflow rate from the disk.
			In a similar manner, we also estimate the angular momentum lose and kinetic power of the disk wind as follows;
				\begin{align}
					d\dot{J}_{\rm wind}(t, r) &= d\dot{M}_{\rm acc}(t, r) \times j(t, r), \label{dot_j_wind}\\
					d\dot{E}_{\rm wind}(t, r) &= d\dot{M}_{\rm acc}(t, r) \times \frac{1}{2} \xi v_{\rm esc}^{2}(t, r), \label{dot_e_wind}
				\end{align}
			where $j(t, r), v_{\rm esc}(t, r), \xi$ are the specific angular momentum, escape velocity, and the fudge factor which absorbs our ignorance of the details of the wind.
			We set $\xi = 0.1$.
			For the disk rotating at the Kepler velocity, $j(t, r) = \sqrt{GM_{\rm BH}(t)r}$.
			We integrate  Eqs.\eqref{dot_j_wind} and \eqref{dot_e_wind} in radial direction.
			The disk state can be described as follows, depending on $r$ and $\dot{M}_{\rm acc}(t, r_{\rm disk})$:
			\noindent [I] pure NDAF phase: \\
			If $\log{\left(\dot{M}_{\rm acc}(t,r_{\rm disk})/M_{\odot})\right)} \geq \log{(r_{\rm disk}/r_{\rm s})} -2.5$, then

				\begin{align}
					\dot{M}_{\rm BH}(t) &= \dot{M}_{\rm acc}(t,r_{\rm disk}) \left( \frac{r_{\rm isco}}{r_{\rm disk}}\right)^{0.4s_{\rm g}}, \label{dotM_bh_ndaf}\\
					\dot{J}_{\rm wind}(t) &= \frac{0.8 s_{\rm g} c^{2}}{0.8s_{\rm g} +1} \frac{j(r_{\rm disk})}{\sqrt{r_{\rm disk}}} \frac{\dot{M}_{\rm acc}(t, r_{\rm disk})}{r_{\rm disk}^{0.4s_{\rm g}}}\left[r^{0.4s_{\rm g} +\frac{1}{2}}\right]_{r_{\rm isco}}^{r_{\rm disk}}, \label{dotj_wind_ndaf}\\
					\dot{E}_{\rm wind}(t) &= \frac{0.02s_{\rm g}c^{2}}{1-0.4s_{\rm g}} \frac{r_{\rm s}}{r_{\rm disk}^{0.4s_{\rm g}}}\dot{M}_{\rm acc}(t,r_{\rm disk})\left[r^{0.4s_{\rm g}-1}\right]_{r_{\rm disk}}^{r_{\rm isco}}. \label{dote_wind_ndaf}
				\end{align}

			\noindent [II] A mixture of the NDAF and ADAF phases:\\
			If $ \log{(r_{\rm isco}/r_{\rm s})} + s_{\rm g}\log{(t,r_{\rm disk} / r_{\rm isco})} -2.5 \leq \log{(\dot{M}_{\rm acc}(r_{\rm disk}) / M_{\odot})} \leq \log{(r_{\rm disk}/r_{\rm s})}-2.5$, then;
				\begin{align}
					\dot{M}_{\rm BH}(t) &= \dot{M}_{\rm acc}(t,r_{\rm disk})\left(\frac{r_{\rm t}}{r_{\rm disk}}\right)^{s_{\rm g}}\left(\frac{r_{\rm isco}}{r_{\rm t}}\right)^{0.4s_{\rm g}}, \label{dotm_bh_ndafadaf}\\
					\dot{J}_{\rm wind}(t) &= \nonumber \\
					&\hspace{-1cm}\frac{2s_{\rm g}}{2s_{\rm g}+1}\sqrt{GM_{\rm BH}r_{\rm disk}}\dot{M}_{\rm acc}(t,r_{\rm disk})\left(1- \left(\frac{r_{\rm t}}{r_{\rm disk}}\right)^{\frac{2s_{\rm g}-1}{2}}\right)\nonumber\\
					 &\hspace{-1cm}+ \frac{0.8s_{\rm g}}{0.8s_{\rm g} + 1} \frac{j(r_{\rm disk})}{\sqrt{r_{\rm disk}}}\frac{\dot{M}_{\rm acc}(t,r_{\rm disk})  }{r_{\rm t}^{0.4s_{\rm g}}} \left( \frac{r_{\rm t}}{r_{\rm disk}}\right)^{s_{\rm g}} \left[r^{0.4s_{\rm g} + \frac{1}{2}}\right]_{r_{\rm isco}}^{r_{\rm t}}, \label{dotj_wind_ndafadaf}\\
					\dot{E}_{\rm wind}(t) &= \frac{0.1s_{\rm g}}{2(1-s_{\rm g})}\frac{\dot{M}_{\rm acc}(t,r_{\rm disk})c^{2}}{(r_{\rm disk}/r_{\rm s})^{s_{\rm g}}}\left[\left( \frac{r}{r_{s}} \right)^{-1+s_{\rm g}}\right]_{r_{\rm disk}}^{r_{\rm t}}\nonumber\\
					&\hspace{-2cm}+ \frac{0.1 \times 0.2 s_{\rm g}c^{2}}{1-0.4s_{\rm g}} \frac{r_{\rm s}}{r_{\rm t}^{0.4s_{\rm g}}}\left( \frac{r_{\rm t}}{r_{\rm disk}}\right)^{s_{\rm g}}\dot{M}_{\rm acc}(t,r_{\rm disk}) \left[ r^{0.4s_{\rm g} -1} \right]_{r_{\rm t}}^{r_{\rm isco}}. \label{dote_ndafadaf}
				\end{align}
			$r_{t}$ is transition radius given by the contour in Figure.3 of \cite{Kohrietal2005}, $r_{\rm t} = r_{\rm s} \left( 10^{2.5} \frac{\dot{M}_{\rm acc}(r_{\rm disk})}{M_{\odot}} \left(\frac{r_{\rm disk}}{r_{\rm s}} \right)^{-s_{\rm g}}\right)$ .

			\noindent [III] pure ADAF:\\
			If $ \log{ (\dot{M}_{\rm acc}(t,r_{\rm disk}) / M_{\odot})} < \log{(r_{\rm isco}/r_{\rm s })} + s_{\rm g}\log{(r_{\rm disk} / r_{\rm isco})} -2.5$, then; 

				\begin{align}
					\dot{M}_{\rm BH}(t) &= \dot{M}_{\rm acc}(t,r_{\rm disk}) \left( \frac{r_{\rm s}}{r_{\rm disk}}\right)^{s_{\rm g}}, \label{dotm_bh_adaf}\\
					\dot{J}_{\rm wind}(t) &= \nonumber \\
					& {\hspace{-1cm}}\frac{2s_{\rm g}}{2s_{\rm g} + 1} \sqrt{GM_{\rm BH} r_{\rm disk}}\dot{M}_{\rm acc}(t,r_{\rm disk}) \left( 1 - \left(\frac{r_{\rm isco}}{r_{\rm disk}}\right)^{\frac{2s_{\rm g}+1}{2}}\right), \label{dotj_wind_adaf}\\
					\dot{E}_{\rm wind}(t) &=  \frac{0.1s_{\rm g}}{2(1-s_{\rm g})} \frac{\dot{M}_{\rm acc}(t,r_{\rm disk})c^{2}}{(r_{\rm disk} /r_{\rm s})^{s_{\rm g}}} \left[ \left(\frac{r_{\rm s}}{r}\right)^{1-s_{\rm g}}\right]_{r_{\rm disk}}^{r_{\rm s}}. \label{dote_wind_adaf}
				\end{align}
			$j(r_{\rm disk})$ is the specific angular momentum at $r_{\rm disk}$.
			These equations are used to describe the evolution of the system as follows:
				\begin{align}
					\dot{M}_{\rm wind}(t) &= \dot{M}_{\rm acc}(t,r_{\rm disk}) -\dot{M}_{\rm BH}(t), \label{dotm_wind}\\
					\dot{J}_{\rm disk}(t) &= \dot{J}_{\rm fb}(t) - \dot{J}_{\rm BH}(t) -\dot{J}_{\rm wind}(t).
				\end{align}

		\subsubsection{Disk wind and supernova explosion}	
			In the previous works, it was hypothesized that all the wind from the disk is able to escape from the system \citep[e.g.,][]{MacFadyenWoosley1999,Kohrietal2005,Kumaretal2008}.
			However, there would be the interaction between the wind and the envelope.
			If the wind can not push the envelope outward, the explosion would not be launched.
			
			So, we consider an outcome of the interaction between the disk wind and the envelope.
			The condition for which the wind can expand outward against the infalling envelope is a balance between the ram pressure of the wind and that of the infalling material \citep{MaedaTominaga2009}.
			If the wind pressure does not exceed the ram pressure of the infalling material, the wind should be pushed back to the disk or the BH.
			In this case, the wind material should indeed be counted as an accretion.
			The infalling material ram pressure $P_{\rm ram}$ and the wind pressure $P_{\rm wind}$ are estimated as follows if we consider a spherically spreading disk wind;

				\begin{align}
					P_{\rm ram}(t) &= \rho_{0,i} \left( \frac{r_{0,i}}{r_{\rm com}}\right)^{\frac{3}{2}} \frac{GM_{\rm BH}(t)}{r_{\rm com}}, \label{P_envelope}\\
					P_{\rm wind}(t) &= \frac{\dot{E}_{\rm wind}(t)}{4 \pi r_{\rm com}^{2}\left(\frac{2GM_{\rm BH}(t)}{r_{\rm com}}\right)^{1/2} }. \label{P_wind}
				\end{align}
			Here we define the competition radius $r_{\rm com}$, which gives a rough measure of the radius at which the wind and the envelope collide and interact with each other.

			While the exact position of $r_{\rm com}$ is unknown, we can estimate this location by considering where the gases from the envelope fall onto the central system and where the wind is mainly blown.
			The gases from the envelope fall onto $r_{{\rm fb}, i}(t)\simeq r_{\rm disk}$.
			The disk wind is likely to be blown along the geometry of the disk.
			In addition, the geometry of the disk would be very thick and the wind would be blown mainly in the outer side of the disk.
			Thus we adopt the competition radius as the disk radius, i.e.,$r_{\rm com} \simeq r_{\rm disk}(t)$.
			Once $P_{\rm wind}(t) \geq P_{\rm ram}(t)$, we assume the explosion is successful and there is no accreting material from the envelope any more after that.
			During the time duration when the relation $P_{\rm wind}(t) \leq P_{\rm ram}(t)$ is met, we assume that the wind falls onto the BH. 
			These situations are described as follows:

			\indent Once $P_{\rm wind}(t) \geq P_{\rm ram}(t)$ is satisfied at some point during the simulation, the SN takes place and the system evolves after the SN as follows: 
				\begin{align}
					\dot{M}_{\rm fb}(t) &= 0, \dot{J}_{\rm fb}(t) = 0 ~~(t_{\rm win} \leq t), \\
					E_{\rm wind}(t) &= \int_{t_{\rm win}}^{t} \dot{E}_{\rm wind}(t^{\prime}) ~dt^{\prime}, ~~M_{\rm wind}(t) = \int_{t_{\rm win}}^{t} \dot{M}_{\rm wind}(t^{\prime}) ~dt^{\prime}.
				\end{align}
			\indent If  $P_{\rm wind}(t) \leq P_{\rm ram}(t)$ is always satisfied until all the progenitor materials collapse to the central system, basically no SN explosion takes place.
			Still, if the disk is left sufficiently massive, it could still energize the disk wind, which can now propagate outward with virtually zero ram pressure inserted by the collapsing envelope.
			This situation is described as follows:
				\begin{align}
					\dot{M}_{{\rm fb},i}(t) &= 0, ~\dot{J}_{{\rm fb},i}(t) = 0,~~ t_{\rm end}\leq t\\
					E_{\rm wind}(t) &= \int_{t_{\rm end}}^{t} \dot{E}_{\rm wind}(t^{\prime}) ~dt^{\prime}, ~~M_{\rm wind}(t) = \int_{t_{\rm end}}^{t} \dot{M}_{\rm wind}(t^{\prime}) ~dt^{\prime}
				\end{align}
			In the above equations, $t_{\rm win}$ is the time in which the disk wind overwhelms the infalling material in the pressure and $t_{\rm end}$ is the time when all the envelope fall onto the central system.

		\subsubsection{GRB jet}
			
			How the GRB jet is launched is still in debate. 
			A widely discussed physical mechanism is the Blandford Znajek  process \citep{BlandfordZnajek1977}.
			In estimating how much energy is converted to the GRB jet, we use a result by a General Relativistic Magneto-Hydrodynamics (GRMHD) collapsar simulation by \cite{McKinney2005}.
			We adopt a fitted formula of the conversion efficiency as a function of the BH spin parameter ($a$) as follows \citep[see also][]{Kumaretal2008}; 
				\begin{align}
					L_{\rm jet}(t) &= \eta_{\rm jet}(a) \dot{M}_{\rm BH}(t) c^{2},\label{l_jet}\\
					\eta_{\rm jet} &= 0.07\left(\frac{a}{1+\sqrt{1-a^{2}}}\right)^{5}.
				\end{align}
			$L_{\rm jet}$ is the GRB jet luminosity.
			While this formula is derived for the ADAF phase, we use the same formula for the NDAF phase.
			In any case, the NDAF phase is too short to substantially contribute to the jet energy, and thus it does not affect our conclusions.
	
			The GRB jet is considered to be collimated in their geometry based on various observations \citep{Piran_2004,Meszaros_2006,Woosley_Bloom_2006}.
			We adopt 5 degree as the jet collimation angle ($\theta_{\rm jet}$). 
			Because the jet is collimated, pressure of the jet, $P_{\rm jet}(t)$, is estimated as follows;
				\begin{align}
					P_{\rm jet}(t) \simeq \frac{L_{\rm jet}(t)}{4\pi cr_{\rm com}^{2}\left(1-\cos{\theta_{\rm jet}}\right)^{2}}.\label{P_jet}
				\end{align}
			Again, $r_{\rm com}$ is uncertain but we adopt $r_{\rm disk}$ to represent an optimistic situation in which the jet could be launched.
			Like the disk wind case, once the jet could exceed the ram pressure, we regard that a GRB jet is successfully launched.
			Unless the jet overcomes the ram pressure, we do not count the GRB energy until the whole envelope falls onto the  central system.

	\subsection{nucleosynthesis}
		
		Typical core collapse SNe produce about $0.1M_{\odot}$ of $^{56}$Ni, as derived by analyzing their light curves.
		However, GRB-SN1998bw-like SNe are more energetic and produce a larger amount of $^{56}$Ni than canonical SNe \citep{Iwamotoetal1998,Drout_etal_2011,Lyman_etal_2016}.
		In the following, we provide rough estimate of the $^{56}$Ni production in our model, by considering temperature, density and entropy.

		\subsubsection{Nucleosynthesis in the disk wind}
			We have the disk wind energy, mass loss rate and velocity from \Eqsref{dotm_wind} and \eqref{dot_e_wind}.
			If the escaping gases are located at $r_{\rm com}$, the mass loss rate and the escape veloctiy are converted to the wind density, $\rho_{\rm wind}(t)$, as follows:
				\begin{align}
					\rho_{\rm wind}(t) &= \frac{\dot{M}_{\rm wind}(t)}{4\pi r_{\rm com}^{2}}\times \sqrt{\frac{r_{\rm com}}{2GM_{\rm BH}(t)}}. \label{rho_wind}
				\end{align} 
			With a reasonable assumption that the thermal energy of the wind is comparable to the kinetic enegy of the wind, we can roughly estimate the temperature of the wind $T_{\rm wind}(t)$ as follows;
				\begin{align}
					T_{\rm wind}(t) &\simeq \left(\frac{G}{a_{\rm rad}}\frac{\rho_{\rm wind}M_{\rm BH}}{r_{\rm com}} \right)^{\frac{1}{4}},
				\end{align}
			where $a_{\rm rad}$ is the radiation constant.
			In addition, the entropy of the wind per volume, per baryon and in the Boltzmann constant unit ($k_{\rm b}$), $s_{\rm wind}$, is estimated as follows;

				\begin{align}
					s_{\rm wind} \simeq  \frac{4a_{\rm rad}m_{\rm p} T_{\rm wind}^{3} }{3 \rho_{\rm wind}k_{\rm b}}. \label{S_wind}
				\end{align}

			Here, $m_{\rm p}$ is the proton mass.
			Characteristic nucleosynthesis products are highly dependent on the temperature and the entropy.
			$^{56}{\rm Ni}$ is produced at $T \gtrsim 5\times 10^{9}{\rm K}$.
			For the higher entropy, the production of $^{56}$Ni is suppressed due to photodisintegration to $^{4}$He \citep[e.g.,][]{Surman_etal_2006, Surman_etal_2011}.
			We estimate the $^{56}{\rm Ni}$ production and $^{4}{\rm He}$ production as follows;

			\indent [a] $T_{\rm wind} < 5 \times 10^{9} {\rm K}$ (no significant burning): \\
				\indent \indent Temperature is not sufficiently high to produce $^{56}$Ni. Progenitor compositions are kept unchanged. 

			\indent [b] $T_{\rm wind} \geq 5 \times 10^{9} {\rm K}$ (explosive burning):\\
				\indent \indent $^{56}{\rm Ni}$ are produced but a part of nuclei would be disintegrated to $\alpha$ particles depending on its entropy. 
				We adopt the following final mass fractions for ($^{56}{\rm Ni} : ^{4}{\rm He}$) depending on the entropy: $(0.1 :0.9)$ for $s_{\rm wind} > 10$, $(0.5:0.5)$ for $1 < s_{\rm wind} < 10$, and $(1.0:0.0)$ for $s_{\rm wind} < 1$.

			Using these expressions, we integrate the masses of the nucleosynthesis products.

		\subsubsection{Nucleosynthesis in the envelope}
			The envelope may also experience explosive nucleosynthesis by a shock wave launched by the disk wind.
			\cite{MaedaTominaga2009} discussed nucleosynthesis in the envelope including this process, and concluded that the production of $^{56}$Ni is highly sensitive not only to the wind properties but also to the timing when the wind is launched. 
			
			With the initial progenitor density structure of 16TI model in \cite{WoosleyHeger2006} and the input energy as determined by \Eqref{dot_e_wind}, we estimate the maximum mass of $^{56}{\rm Ni}$ produced through explosive nucleosynthesis in the envelope.
			For given wind energy $E_{\rm wind}$, the region in which $T \gtrsim 5\times 10^{9}{\rm K}$ for the efficient $^{56}{\rm Ni}$ production is estimated as follows \citep{Woosley_Weaver_1995}; 
				\begin{align}
					r_{\rm 5E9}(t) = \left(\frac{E_{\rm wind}(t)}{a_{\rm rad}(5\times 10^{9})^{4}} \right)^{\frac{1}{3}}.
				\end{align}

			By applying this prescription to our model, we conclude that the product of $^{56}$Ni is largely negligible for the situations considered in this paper. 
			In our typical models, the wind is launched when the BH becomes already massive; therefore  the amount of the envelope materials available for the explosive nucleosynthesis is already small \citep{MaedaTominaga2009}

			\section{Result: Quasi spherical case}
	
	In our standard model, we adopt the progenitor model 16TI from \cite{WoosleyHeger2006}, with the total angular momentum of the progenitor star set at $J_{\rm standard} \simeq 3.4\times 10^{51}[{\rm cm^{2}}\cdot{\rm s^{-1}}]$.
	The angular velocity distribution is assumed to follow a power law in radius with the index of $\sim -0.19$.
	This set up is very similar to that investigated by \cite{Kumaretal2008}.
	In this section, we assume the wind is blown off in a spherically symmetric manner.

	In $\S 3.1$, we show the results of the standard model.
	In $\S 3.2$, we show the dependence of the outcome on the total angular momentum in the progenitor star.
	In $\S 3.3$, we discuss how the wind parameter $s_{\rm g}$ affects the outcome.

	\subsection{Standard case}

\begin{figure*}
				\centering
				\includegraphics*[width = \textwidth, bb = 0 0 540 378]{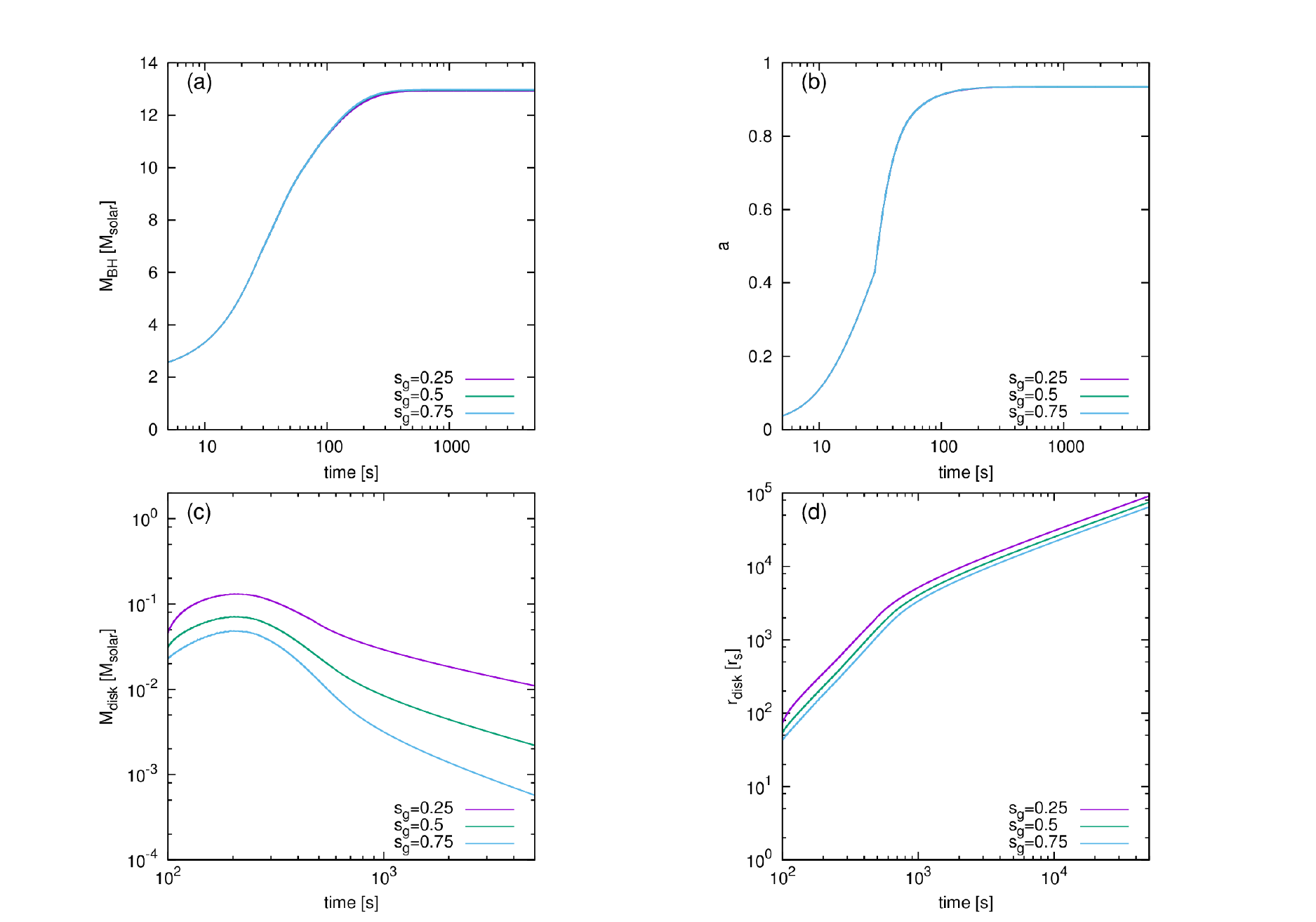}
				\caption{Evolution of the BH and the disk in the quasi-spherical wind case. 
				(a) The BH mass, (b) the BH spin parameter, (c) the disk mass, and (d) the disk radius.
				The results with $s_{\rm g} = 0.25$, $0.5$, and $0.75$ are shown by violet, green, and blue, respectively.}
				\label{standard_evo}
			\end{figure*}

		The evolutions of the BH and the disk in our standard run are shown in \figsref{standard_evo}.
		The final mass of the BH reaches nearly the progenitor mass irrespective of $s_{\rm g}$.
		Similarly, the spin of the BH reaches $\sim 0.9$ regardless of the parameter $s_{\rm g}$ (\figref{standard_evo}b); the wind can not exceed the pressure of the infalling material and only a small fraction of the envelope can be blown by the wind (see below for further details).

		Properties of the disk are shown in \figsref{standard_evo}c and 2d.
		Almost all of the disk materials accrete on the BH and the mass stored on the disk is at most $M_{\rm disk} \sim 0.1 M_{\odot}$. 
		For larger $s_{\rm g}$, the stronger wind is initiated extracting the disk angular momentum more effectively.
		As a result, the radius of the disk is smaller for larger $s_{\rm g}$.
		The evolution of the disk is also faster for larger $s_{\rm g}$.

\begin{figure}
				\centering
				\includegraphics*[width = \columnwidth, bb = 10 0 210 151]{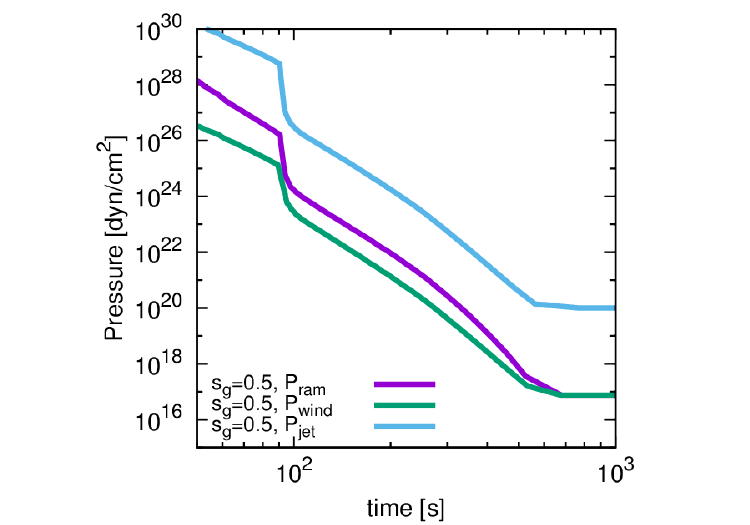}
				\caption{The competition of the ram pressure between the jet, the wind and the infalling envelope, for the quasi-spherical wind case. 
				Shown here are the evolution of the pressure by the infalling envelope (violet), the disk wind (green), and the jet (blue). 
				The constant behavior after $500{\rm s}$ is an artifact.}
				\label{P(base)}
			\end{figure}

		We show the pressure evolution in \figref{P(base)}.
		As mentioned before, the wind cannot exceed the ram pressure of the infalling material.
		This wind power is ultimately a fraction of the gravitation binding energy of the accreting material.
		If the time scale of converting the gravitational energy (i.e., the viscous time scale of the disk) is negligible as compared to the time scale in which the ram pressure of the infalling material decreases significantly (i.e., the time scale of the free fall time of the envelope), there is no way that the wind can overwhelm the infalling envelope, from the consideration of energetics.
		This turns out to be the case for the situation considered here.

\begin{figure*}
				\centering
				\includegraphics*[width  = \textwidth, bb =  30 0 755 201]{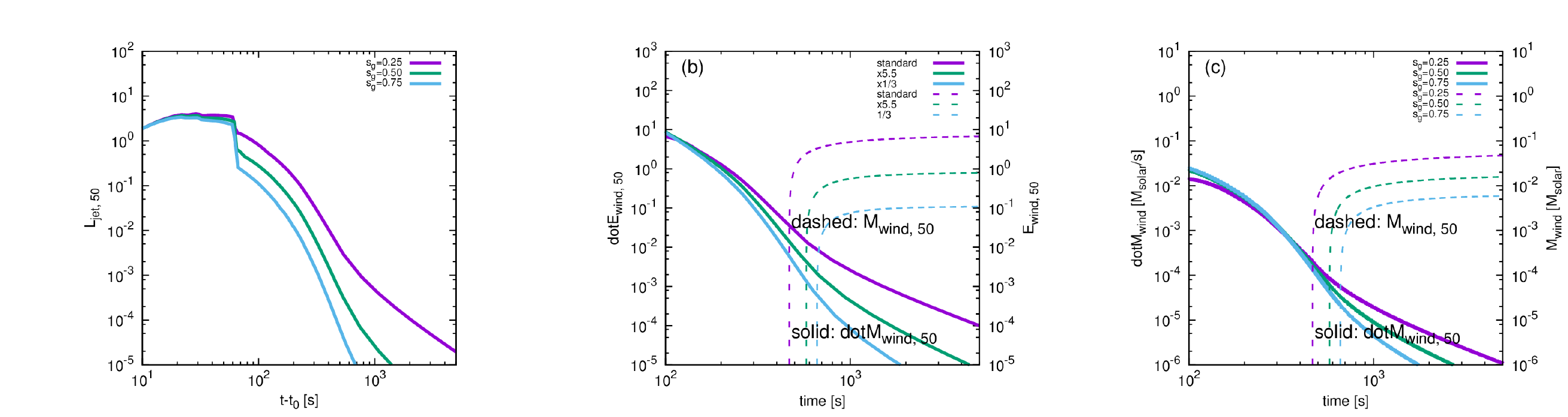}
				\caption{Properties of the jet and the wind with different wind parameter $s_{\rm g} = 0.25$, $0.5$, and $0.75$ (violet, green, and blue, respectively). 
				The panel (a) shows the jet luminosity in $10^{50}$ erg s$^{-1}$. 
				The panel (b) shows the disk wind power (solid; in $10^{50}$ erg s$^{-1}$) and integrated energy reached after the launch of the SN (dashed; in $10^{50}$ erg).
				The panel (c) shows the wind mass loss rate ($M_{\odot}$ s$^{-1}$) and integrated wind mass ($M_{\odot}$).
				}
				\label{Jet and wind}
			\end{figure*}

		\figref{Jet and wind} shows the luminosity of the jet, the kinetic luminosity of the wind, and the mass loss rate by the wind.
		The luminosity of the jet is consistent with the previous study by \cite{Kumaretal2008}.
		However, the predicted properties of the wind are very different, due to the additional consideration on the launch condition. 
		The energy and the mass of the wind in our model are smaller than those of the previous works by one or two orders of magnitudes.
		We conclude that this standard model with a (quasi) spherical wind would not produce an energetic SN.

	\subsection{Dependence on angular momentum}

		In this section, we investigate how the total angular momentum in the progenitor star impacts the outcome.
		While keeping the density profile of 16TI model, we change the angular momentum artificially by multiplying a constant value to the angular momentum distribution in our standard model. 
		In \S 3.2, we fix $s_{\rm g} = 0.5$.

\begin{figure}
				\centering
				\includegraphics*[width = \columnwidth, bb = 0 40 360 202]{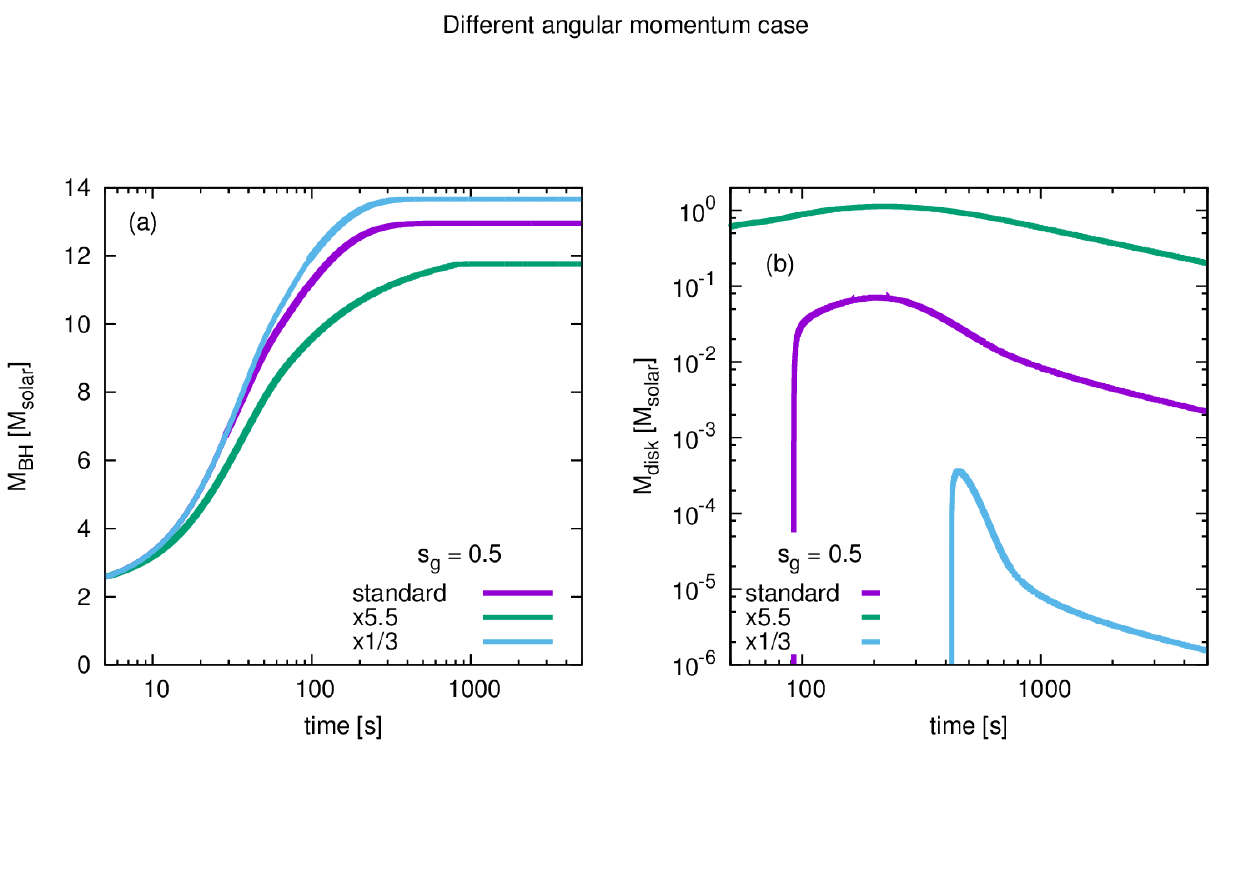}
				\caption{The evolution of (a) the BH and (b) the disk with different angular momentum of the progenitor. 
				Our standard model ($J_{\rm standard}$) is shown by violet, while the models with larger/smaller angular momentum (5.5 or 1/3 $\times J_{\rm standard}$) are shown by green and blue.
				See the caption of Figure 1.}
				\label{property_hikaku}
			\end{figure}

		\figref{property_hikaku} shows the evolution of the BH and disk masses. 
		For larger angular momentum, the material from the envelope more efficiently accretes onto the disk.
		The final mass of the BH is still large ($\simeq 12M_{\odot}$) even though the disk is larger and the wind is stronger than those in the standard model.
		On the other hand, if we reduce the angular momentum by a factor of three, almost all of the progenitor falls onto the BH directly.

\begin{figure}
				\centering
				\includegraphics*[ width = \columnwidth, bb = 10 0 200 151]{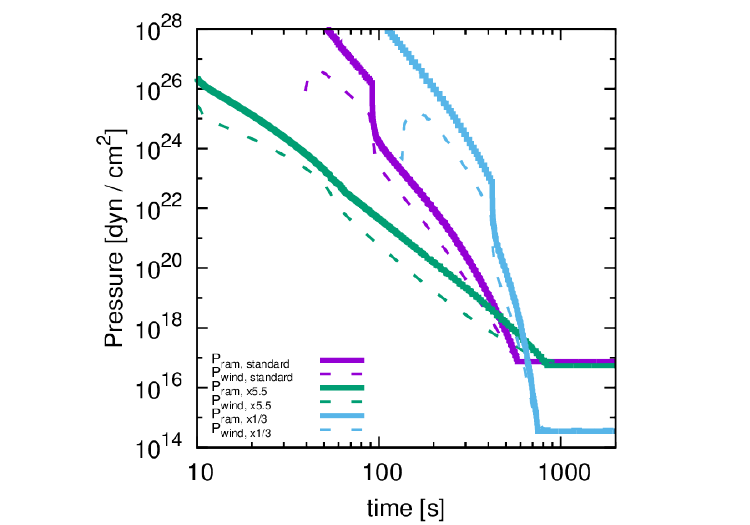}
				\caption{Same as Figure 3, but for different angular momentum (see the caption of Figure 5).}
				\label{p_hikaku}
			\end{figure}

		\figref{p_hikaku} shows the evolution of the wind and ram pressure.
		The pressure of the infalling material is still large, always exceeding the wind pressure.

\begin{figure*}
				\centering
				\includegraphics*[width = \textwidth,  bb = 30 0 755 201]{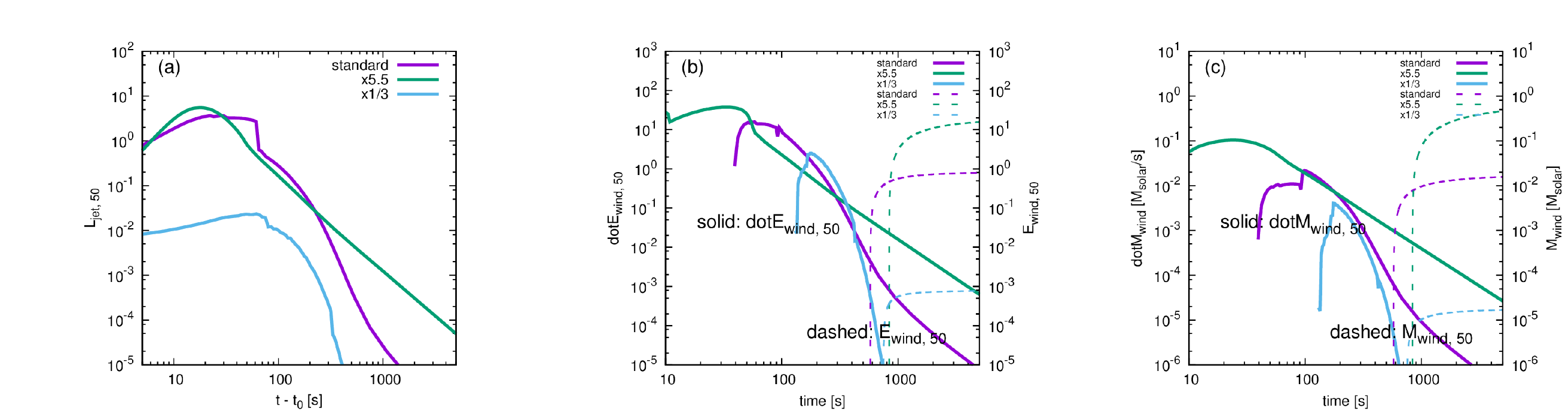}
				\caption{Same as Figure 4, but for different angular momentum (see the captions of Figures 4 and 5).}
				\label{jet_wind_property_hikaku}
			\end{figure*}

		\figref{jet_wind_property_hikaku} show the jet luminosity and the properties of the wind.
		The jet luminosity is sensitive to the angular momentum.
		For larger angular momentum, the spin of the BH reaches to the maximum ($a > 0.9$) more rapidly.
		The jet luminosity then exceeds the typical GRBs ($L_{\rm jet} \simeq 10^{51}{\rm erg/s}$).
		In addition, the disk evolution is slower than the standard model, resulting in a more slowly declining jet power.
		On the other hand, for the smaller angular momentum case, the BH spin is slow in the first stage.
		As the disk accretion proceeds, the spin of the BH becomes faster while the accretion rate becomes smaller.
		Due to a combination of these effects, the luminosity of the jet is nearly constant in the first $\sim 200 {\rm s}$.

		The result is similar to the standard model on the capability of the wind to launch an SN.
		For the large angular momentum case, the energy and the mass of the ejecta are comparable to the canonical SN.
		However, even for the most rapidly rotating progenitor, the resulting SNe are not as energetic as the observed GRB-SNe.

	\subsection{Model survey}

\begin{figure*}
				\centering
				\includegraphics*[bb = 40 216 985 542, width = \textwidth]{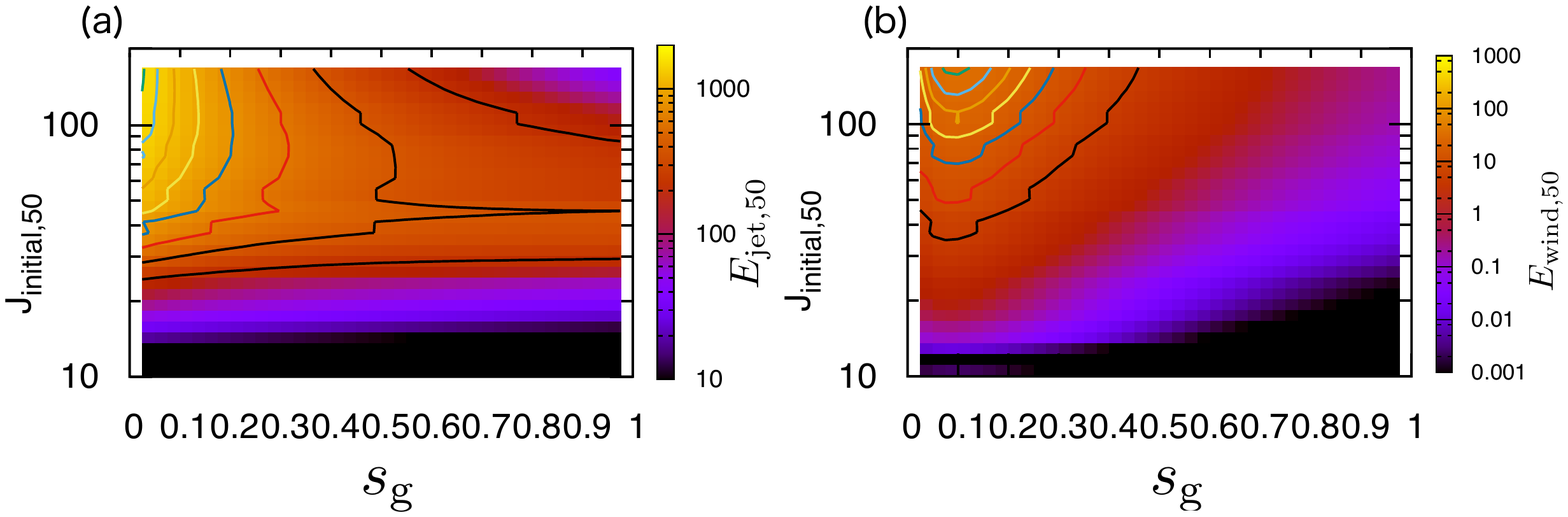}
				\caption{The Jet energy (a) and the wind energy (b) (in $10^{50}{\rm erg}$) as functions of the wind parameter $s_{\rm g}$ (x-axis) and total angular momentum of the progenitor (in $10^{50}{\rm g\cdot cm^{2}\cdot s^{-1}}$, y-axis).}
				\label{wind_jet_js_dependence}
			\end{figure*}

		In addition to the angular momentum examined in \S 3.2, we study the effect of the wind parameter in this section. We vary $s_{\rm g}$ in the range $0 < s_{\rm g} < 1$. 
		\figref{wind_jet_js_dependence} shows the jet energy and the wind energy as a function of the angular momentum and $s_{\rm g}$.
		For larger $s_{\rm g}$, the angular momentum taken by the wind is larger.
		Therefore, the wind power decreases more quickly, leading to smaller wind energy. 

\begin{figure}
				\centering
				\includegraphics*[width = \columnwidth, bb = 10 0 270 201]{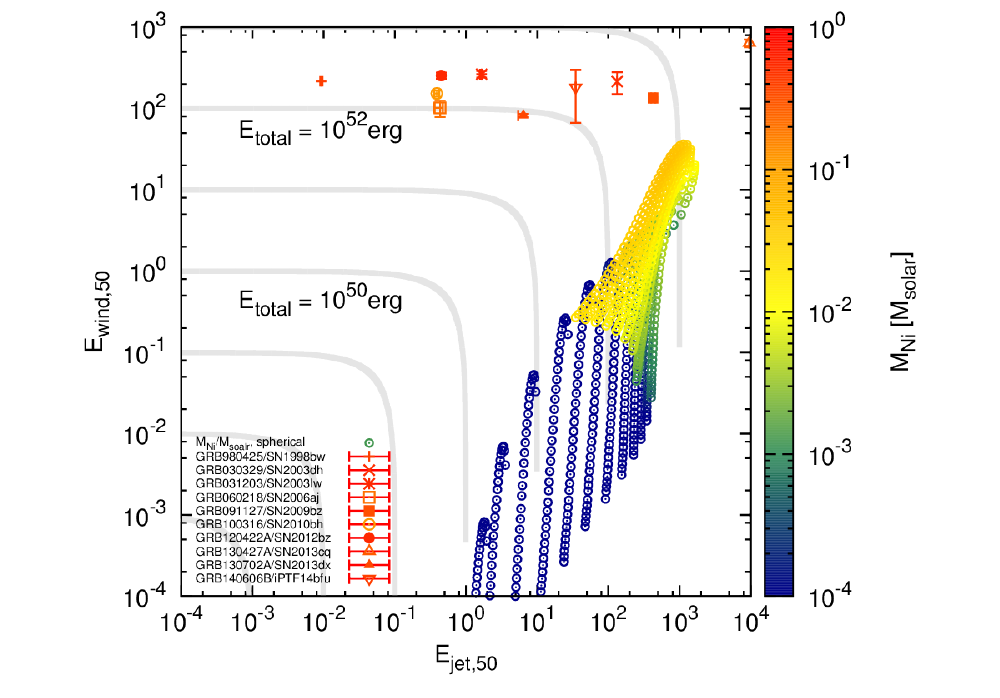}
				\caption{The expected relation between the jet energy and wind energy (in $10^{50}{\rm erg}$). 
				Also shown is the ejected mass of $^{56}$Ni as shown by different color (in $M_{\odot}$). 
				The observationally GRB-SNe properties are also shown \citep[red points;][]{Toy_etal_2016}.}
				\label{Ni_mass}
			\end{figure}

		The resulting \Ni mass, the jet energy, and the wind energy are shown in \figref{Ni_mass}.
		The energy of the jet ranges in $10^{50-53}{\rm erg}$, covering the properties of the observed GRBs.
		In our models examined here, the wind alone does not account for the energy of the GRB-SNe.
		Further, the predicted amount of \Ni is also smaller than the typical value derived for the observed GRB-SNe.

\begin{figure}
				\centering
				\includegraphics*[width = \columnwidth, bb = 10 0 200 151]{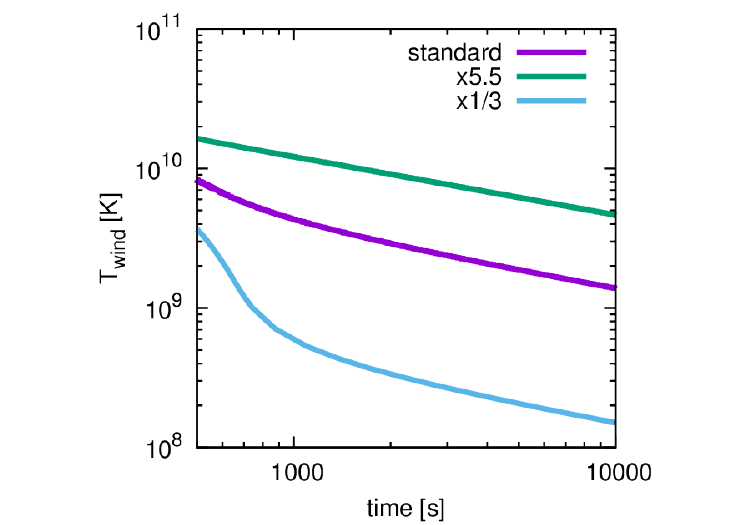}
				\caption{Evolution of the temperature at the launch of the disk wind, with different angular momentum of the progenitor. We fix $s_{\rm g} = 0.5$ in this figure.}
				\label{temple}
			\end{figure}

		\figref{temple} shows the evolution of the temperature of the wind at its base.
		Only the most rapidly rotating progenitor model can have the wind whose temperature exceeds $5\times 10^{9}{\rm K}$ for sufficiently long duration to effectively produce \Ni.

		For example, our model predicts a more energetic SN associated with a more energetic GRB, which would not explain GRB-SNe associated with weak GRBs. 
		Furthermore, the energy of the wind is smaller than those derived for GRB-SNe, and the produced \Ni mass is not enough to be compatible to the observationally derived values.
				
		From the analyses in \S 3, we conclude that it is not possible to construct a self-consistent model for GRBs and GRB-SNe in the disk wind-driven collapsar model, as long as the quasi-spherical wind is assumed.

\section{Collimated disk wind case}
	To remedy the problem in launching the wind against the ram pressure, one possibility is to consider collimation of the wind.
	We take into account the collimation by considering the enhancement of the wind pressure artificially.
	We change \Eqref{P_wind} as follows;
		\begin{align}
				P_{\rm wind,colli} &= \frac{\dot{E}_{\rm wind}(t)}{4\pi r_{\rm com}^2\left(1-\cos{\theta_{\rm wind}}\right)^{2}\left(\frac{2GM_{\rm BH}(t)}{r_{\rm com}}\right)^{1/2}},\label{collimated_pressure}
		\end{align}

	where $\theta_{\rm wind}$ is the collimation angle of the disk wind.

	If we adopt the collimated wind, there would be two different flow patterns to consider. 
	In the following, we consider two different configurations. 
	In Case OUT (figure 1, c-I), we consider that the accretion is stopped after the launch of the SN explosion by the disk wind. 
	Even if the base of the wind is collimated, the lateral expansion may rapidly cover a large solid angle to prevent further accretion. 
	This is similar to the quasi-spherical wind case discussed in \S 3. 
	In addition, we consider Case IN-OUT (Figure 1, c-II), in which we assume that the accretion from the envelope continues even after the launch of the SN explosion.  

\begin{figure}
				\centering
				\includegraphics*[bb = 10 0 200 151, width = \columnwidth]{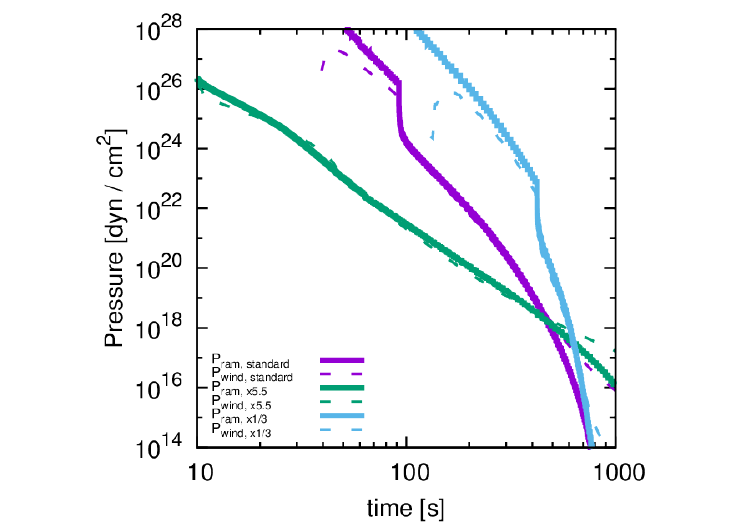}
				\caption{The evolution of pressure for Case IN and Case In-OUT (note this property is the same for both cases). 
				The opening angle of the wind at the base is taken as $\theta_{\rm wind} = 30^{\circ}$. 
				The other parameters are same as shown in \figref{p_hikaku}}
				\label{colli_press}
			\end{figure}

	\figref{colli_press} shows the pressure evolution for a model in which we fix $\theta_{\rm wind} = 30^{\circ}$, and $s_{\rm g} = 0.5$.
	Owing to the geometry of the wind, the wind pressure is larger than that of the spherical case by factor of a few, and the wind launches an SN earlier.
	Therefore, the energy and the mass of the wind-driven SN are larger.

\begin{figure*}
				\centering
				\includegraphics*[bb = 0 0 540 378, width = \textwidth]{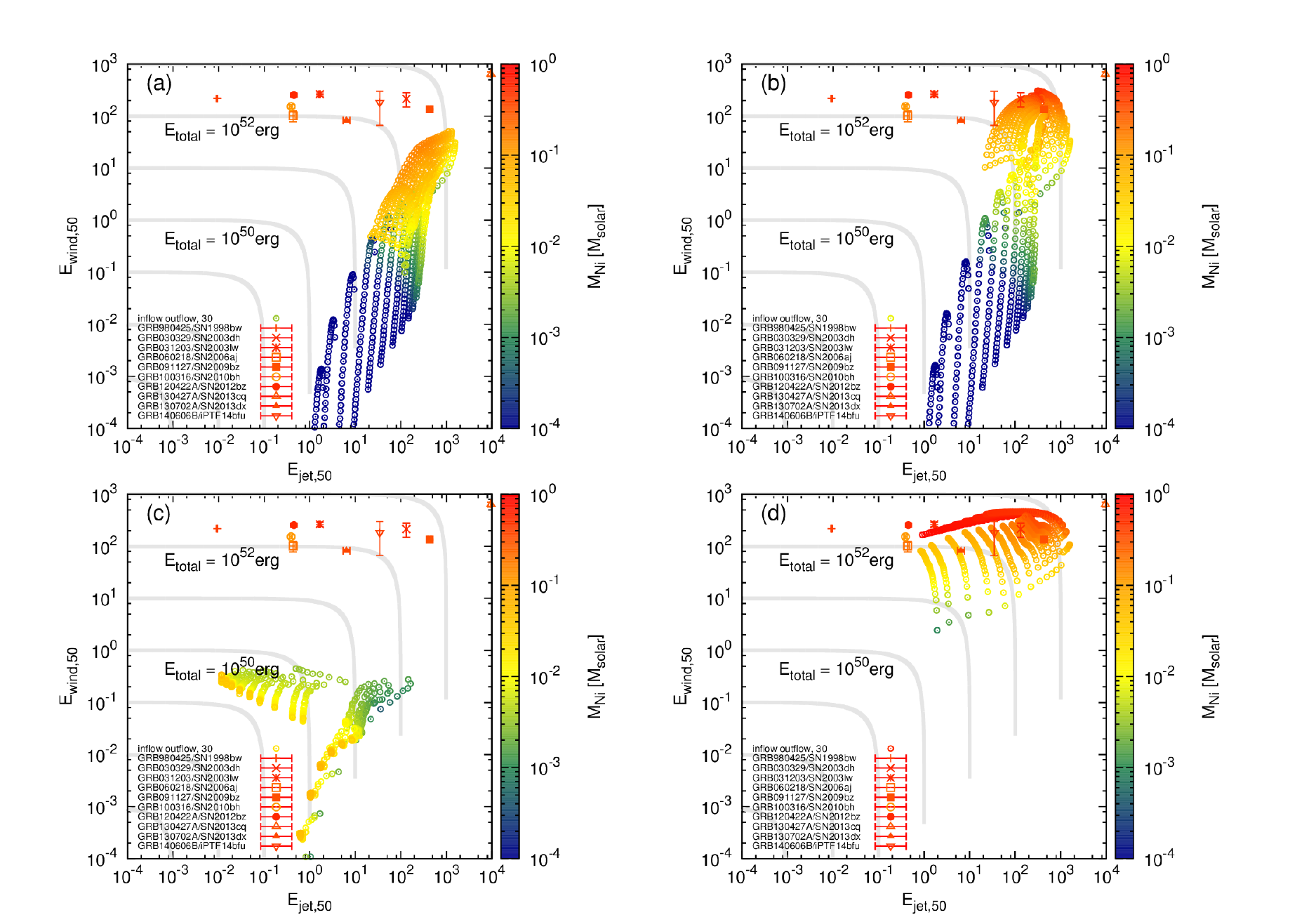}
				\caption{The jet energy and the wind energy (in $10^{50}{\rm erg}$), together with $M$($^{56}$Ni) (in $M_{\odot}$) for different cases of the wind collimation. 
				(a) Case OUT, $\theta_{\rm wind} = 30^{\circ}$,(b) Case IN-OUT, $\theta_{\rm wind} = 30^{\circ}$. 
				(c) Case OUT, $\theta_{\rm wind} = 10^{\circ}$,and (d) Case IN-OUT, $\theta_{\rm wind} = 10^{\circ}$. Observational properties (red points) are taken from Table.6 from \cite{Toy_etal_2016}.
				}
				\label{colli_jet_wind_ni}
			\end{figure*}

	\figref{colli_jet_wind_ni} shows the results of the same progenitors shown in \S 3.1 with different $\theta_{\rm wind}$, for Case OUT and Case IN-OUT.
	If we adopt $\theta_{\rm wind} = 30^{\circ}$ for Case OUT, the results are similar to the spherical case.
	While if we adopt $\theta_{\rm wind} = 30^{\circ}$ for Case IN-OUT, the central engine is active for long time duration thanks to the continuous accretion, therefore the energies of the wind and jet are both enhanced.
	The models with the largest angular momentum result in the energetic jet and wind-driven SNe, which are consistent with the properties of the observed GRB-SNe.

	The predicted relations are very different if we adopt $\theta_{\rm wind} = 10^{\circ}$.
	For case OUT, the wind pressure is sufficiently high to launch the SN before the disk has evolved substantially.
	Lacking the further accretion power in this case, the energies of the wind and the jet are smaller by orders of magnitude than Case IN-OUT.
	On the other hand, if we adopt Case IN-OUT, the inflow can feed the gases to the central system and provides a continuous power.

	These results show that two conditions should be satisfied in order to explain the observed properties of (some) GRB-SNe within the context of the collapsar model: (1) the wind should be collimated to launch the wind-driven explosion sufficiently early, but (2) the wind-driven shock wave would not significantly expand laterally so that continuous accretion can energize the wind.
	Only the rapid rotating models ($J \gtrsim 10^{52} {\rm [cm^{2}\cdot s^{-1}]}$) can produce the wind with a sufficiently large energy to be compatible to the kinetic energy of typical GRB-SNe.

			\section{Progenitor dependence}
	In \S 4, we have concluded that the collimated wind case can be a candidate to explain some GRB-SNe, especially energetic events like GRB030329/SN2003dh or GRB091127/SN2009bz.
	However, the SNe associated with low-luminosity GRBs (e.g., GRB980425/SN1998bw) are not readily explained by our models.

	To investigate whether this is remedied by introducing the diversity related to the progenitor stars, we further perform the simulations based on additional three progenitor models with different masses ($10.7, 21.1, 27.9 M_{\odot}$).
	Note that our standard model, 16TI, has the pre-collapse mass of $13.8M_{\odot}$.
	The total angular momentum is again adopted as a parameter, which is varied from the nearly breaking-up case to the nearly whole collapse.
	We adopt both of Case OUT and Case IN-OUT.

\begin{figure*}
				\centering
				\includegraphics*[ width = \textwidth, bb = 0 0 540 378]{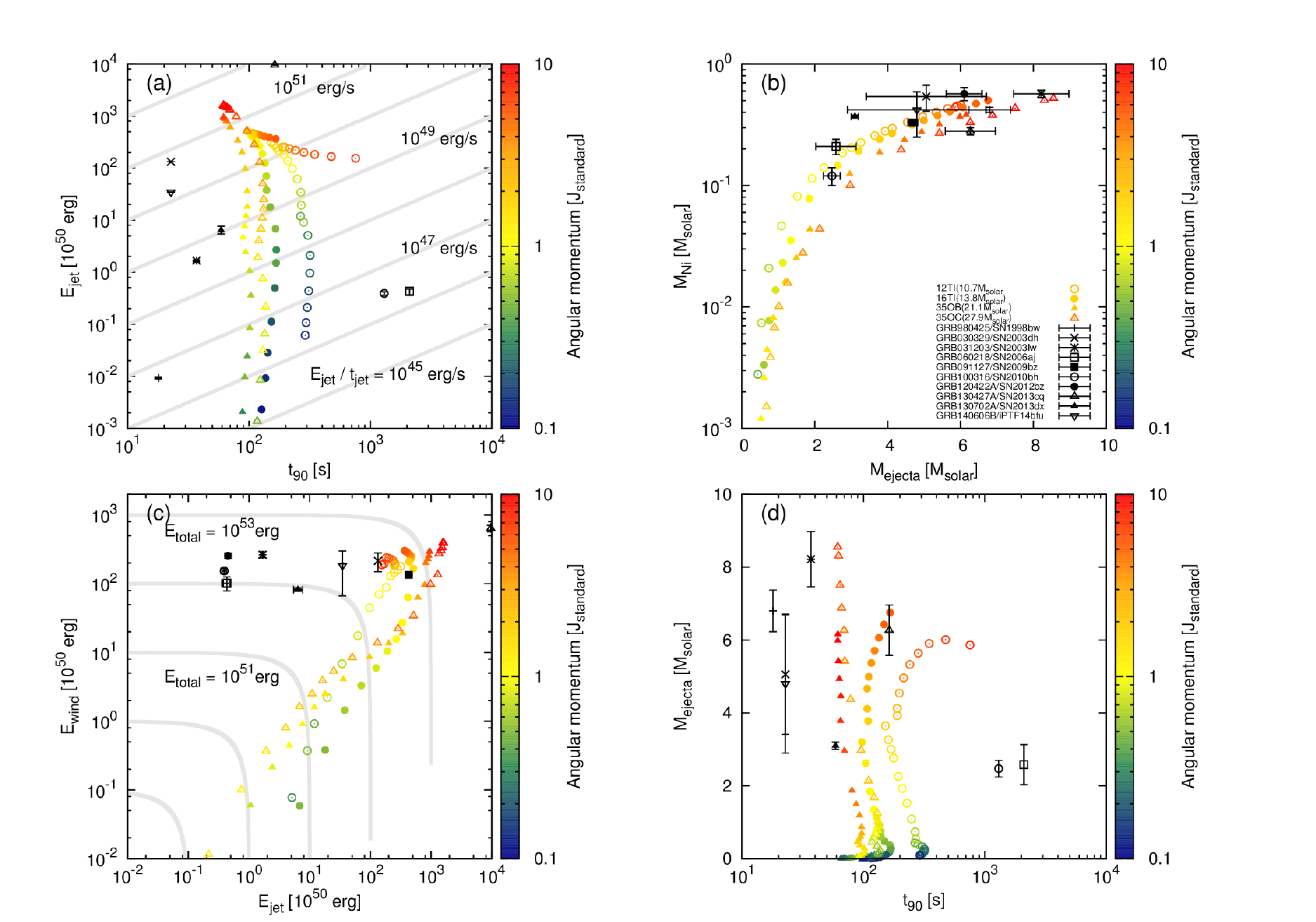}
				\caption{Properties of the jet and wind-driven SNe, for Case In-OUT, $\theta_{\rm wind} = 30^{\circ}$, and $s_{\rm g} =0.3$. 
				Different symbols are used for progenitor models with different masses. 
				Different colors are used for different angular momentum of the progenitor. 
				(a) Jet (GRB) properties, (b) SN ejecta properties, (c) energetics of the jet and the SNe, and (d) GRB and SN properties.
				Observationally derived properties of GRBs and GRB-SNe are from \cite{Toy_etal_2016}.}
				\label{multi_prop_s03}
			\end{figure*}

\begin{figure*}
				\centering
				\includegraphics*[ width = \textwidth, bb = 0 0 540 378]{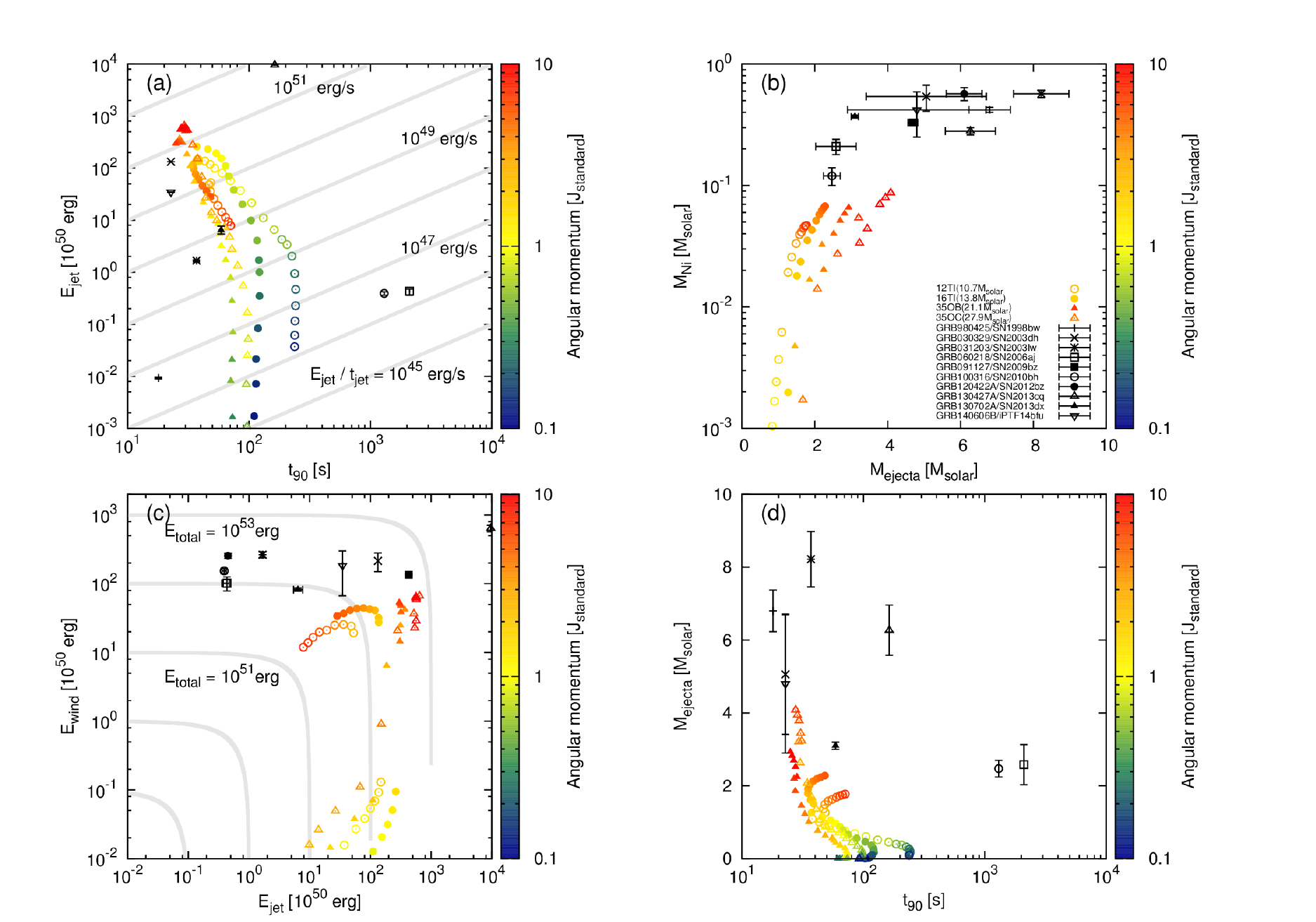}
				\caption{Same as Figure 13 but for $s_{\rm g} = 0.9$.}
				\label{multi_prop_s09}
			\end{figure*}

	\figsref{multi_prop_s03} and \ref{multi_prop_s09} show the results from these models.
	We fix $s_{\rm g} = 0.3, 0.9$ in \figsref{multi_prop_s03} and \figsref{multi_prop_s09}, respectively.
	We find that the outcome is sensitive to the angular momentum, but not to the mass of the progenitor star.
	\figsref{multi_prop_s03}(a) and \ref{multi_prop_s09}(a) show the jet properties, the duration time and the jet energy ($t_{90}$ and $E_{\rm jet}$).
	We could construct models consistent with GRB-SNe in the jet properties, except for ultra-long GRBs, if we adopt $s_{\rm g} \simeq 0.9$.
	For $s_{\rm g}=0.3$, the jet durations are longer than the observed ones by a factor of a few to an order of magnitude.
	Our models cover a range of $E_{\rm jet}$ ($10^{47}-10^{53} ~{\rm erg}$) and $E_{\rm jet}/t_{\rm jet}$ ($10^{45}-10^{52}{\rm erg \cdot s^{-1}}$) by varying the angular momentum.
	The model thus covers the observer diversity in the jet properties.

	\figsref{multi_prop_s03}(b) and \ref{multi_prop_s09}(b) show the properties of SNe, the ejected mass and produced \Ni mass ($M_{\rm ejecta}$ and $M_{\rm Ni}$).
	Contrarily to the jet properties, the results are most consistent with observationally inferred properties for $s_{\rm g} \simeq 0.3$.
	Interestingly, our models lead to a positive correlation between the ejecta mass and $^{56}{\rm Ni}$ mass, which follows the relation derived from the observed GRB-SNe samples \citep{Cano_2013, Lyman_etal_2016, Toy_etal_2016}.

	\figsref{multi_prop_s03}(c) and \ref{multi_prop_s09}(c) show the relation between the jet energy and the wind energy ($E_{\rm jet}$ and $E_{\rm wind}$).
	Again, the properties are mainly controlled by the angular momentum rather than the progenitor mass.
	Our models can explain only the most energetic events like GRB030329/SN2003dh and GRB091127/SN2009bz if we take $s_{\rm g} \simeq 0.3$.
	If we consider only the sum of the jet and wind energies (see \S 6 for the possibility to convert the jet energy to the SN energy),
	our models cover the range of the values observed for GRB-SNe.
	One model fails to explain the properties of the ultra-long GRBs.
	The total jet energy can be explained, but the long duration is never reproduced.
	This suggests that the typical compact progenitor models for GRBs would not apply to the ultra-long GRBs \citep{Kashiyama_etal_2013,Nakauchi_etal_2013,Ioka_etal_2016}.

	\figsref{multi_prop_s03}(d) and \ref{multi_prop_s09}(d) show the jet duration time and the ejected mass ($t_{\rm 90}$ and $M_{\rm ejecta}$).
	Our results show a negative correlation between $t_{\rm 90}$ and $M_{\rm ejecta}$, which might also be seen in the observation.
	For smaller $s_{\rm g}$, the model explains a range of $M_{\rm ejecta}$ as observed, but the duration of the jet is too long to compare to the observations.
	On the other hand, for larger $s_{\rm g}$, the range of $M_{\rm ejecta}$ is smaller but $t_{\rm 90}$ is within the observations except ultra-long GRB-SNe.

	We conclude that some observed properties are explained by our model.
	The diversity in the properties of the jet and SN is mainly driven by different angular momentum in this model.
	The progenitor mass does not affect the results sensitively.
	The models most consistent with the properties of the GRB-SNe are found for $s_{\rm g} \simeq 0.3$.
	The same models fail to fully explain the properties of the jet in a self-consistent way.
	This is further discussed in \S 6.

\section{Discussion}
	\subsection{Expected observational counterparts}

		The outcome of our model is sensitive to the angular momentum of the progenitor star, and the best models to account for the GRB-SNe are found for the largest angular momentum.
		This might indicate that we would expect a variety of observational counterparts for the collapsar scenario, not only the GRB-SNe, depending on the angular momentum of the progenitor star.
		The expected observational counterparts by our models are listed in \tabref{GRB-SNe03} ($s_{\rm g} \simeq 0.3$) and \ref{GRB-SNe09} ($s_{\rm g} \simeq 0.9$).
		Rapidly rotating progenitors would produce energetic GRB-SNe, like GRB030329/SN2003bz and GRB091127/SN2009bz.
		On the other hand, slowly rotating progenitors would lead to less energetic events like low-luminosity GRBs without SNe \citep{Dexter_Kasen_2013, Kashiyama_Quataert_2015}.

		To explain the observed properties of GRBs and GRB-SNe, we find that even larger angular momentum is required than postulated by the previous works \citep{Kohrietal2005,Kumaretal2008}.
		Indeed, our standard model with the angular momentum similar to that assumed by \cite{Kumaretal2008} results in canonical SN energy.
		
			\begin{table*}							
				\caption{Expected counterparts (Case IN-OUT with $s_{\rm g} = 0.3$ and $\theta_{\rm wind} = 30^{\circ}$).}
				\label{GRB-SNe03}
				\centering
				\begin{tabular}{lccc} \toprule
					progenitor & low $J_{\rm initial}$ & standard & high $J_{\rm initial}$ \\ \midrule
					12TI & 1 & 2, 2+A & 3?, 3+A, 3+B \\
					16TI & 1 & 2, 2+A & 3?, 3+A, 3+B \\
					35OB & 1 & 2 & 3?, 3+A, 3+B\\
					35OC & 1 & 2 & 3?, 3+A, 3+B\\
					\bottomrule
					1:ll-GRB, 2:INT-GRB, &3:GRB, A:sub luminous-SNe, &B:hypernovae
				\end{tabular}\\
				ll-GRB:$L_{\rm jet} \leq 10^{48.5}{\rm erg\cdot s^{-1}}$, INT-GRB:$L_{\rm jet} = 10^{48.5-49.5}{\rm erg\cdot s^{-1}}$, GRB:$L_{\rm jet} \geq 10^{49.5}{\rm erg\cdot s^{-1}}$ 
			\end{table*}

			\begin{table*}							
				\caption{Expected counterparts (Case IN-OUT with $s_{\rm g} = 0.9$ and $\theta_{\rm wind} = 30^{\circ}$).}
				\label{GRB-SNe09}
				\centering
				\begin{tabular}{lccc} \toprule
					progenitor & low $J_{\rm initial}$ & standard & high $J_{\rm initial}$ \\ \midrule
					12TI & 1 & 2 & 3, 3+A \\
					16TI & 1 & 2 & 3, 3+A \\
					35OB & 1 & 2 & 3, 3+A \\
					35OC & 1 & 2 & 3, 3+A \\
					\bottomrule
					1:ll-GRB, 2:INT-GRB, &3:GRB, A:sub luminous-SNe, &B:hypernovae
				\end{tabular}\\
				ll-GRB:$L_{\rm jet} \leq 10^{48.5}{\rm erg\cdot s^{-1}}$, INT-GRB:$L_{\rm jet} = 10^{48.5-49.5}{\rm erg\cdot s^{-1}}$, GRB:$L_{\rm jet} \geq 10^{49.5}{\rm erg\cdot s^{-1}}$ 
			\end{table*}

	\subsection{open questions and future perspectives}

		To explain the properties of the observed GRB-SNe, we find that several conditions must be met in the launch of the SN by the disk wind and its subsequent propagation of the envelope.
		(1) The wind must be collimated at its launch, and (2) the wind (and the jet) must not prevent continuous accretion from the envelope to the accretion disk.
		Under these conditions, (only) highly rotating progenitors result in the SN ejecta mass as large as observationally derived.
		The angular momentum of the star in this situation is larger than previously postulated, and an issue is if such stars with the extreme condition could exist in nature.

		While the most rapidly rotating model can explain the global properties of the SN ejecta by the wind-driven explosion, there is no guarantee that the model gives a fully consistent picture.
		In our models, indeed the SN ejecta are dominated by the wind, with only a minor fraction originated in the envelope.
		The ejected \Ni is also dominated by the disk wind component.
		Given the decreasing power of the wind as a function of the time, we would expect that the iron group elements are first blown off from the system (thus in the outermost ejecta), followed by the intermediate mass elements (thus deeper in the ejecta).
		This trend is the opposite to a standard SN model.
		Indeed, there are observational indications that the degree of `mixing' is higher for GRB-SNe than in canonical SNe \citep[e.g.,][]{Maeda_Nomoto_2003,Maeda_etal_2003} with an extreme example of GRB161219B/SN2016jca \citep{Ashall_etal_2017}.
		Still, it is not clear as for whether the extreme element distribution expected from the wind-driven model is compatible to the observed properties of GRB-SNe.

		Geometry of the SN ejecta is another issue.
		The ejecta of observed GRB-SNe are suggested to be quasi spherical rather than extremely bipolar configuration \citep{Maedaetal2008}.
		In our model, or the disk wind-driven SN model in the context of the collapsar scenario in general, both the collimated disk wind and continuous accretion are required to explain the energetics and the ejecta mass of GRB-SNe.
		It is then expected that the resulting explosion is highly asymmetric, more like a jet rather than the quasi-spherical outflow.
		Further addressing this question will require detail hydrodynamic simulations on the interaction between the wind and the envelope in a large scale, covering the whole progenitor star in a computational domain.

		Another possibility is to consider a different central engine.
		The drawback of the quasi-spherical wind is that the viscous time scale, which determines a time lag in covering the gravitational energy to the outflow energy, is too short.
		A potential candidate is a delayed energy input from a rapidly spinning magnetar (highly-magnetized neutron star), instead of a BH.
		Indeed, a magnetar driven explosion has been suggested as an alternative to a BH formation for GRBs and some peculiar SNe \citep{Wheeler_etal_2000,Thompson_2004, Maeda_etal_2007, Maeda_etal_2007_b,Kasen_2010, Woosley_2010,Metzger_etal_2011, Metzger_etal_2014, Cano_etal_2014, Mazzali_etal_2014}, or for some GRB-SNe (GRB060218/SN2006aj: \citealp{Mazzali_etal_2006,Maeda_etal_2007}, ultra-long GRB111209a/SN2011kl; \citealp{Greiner_etal_2015}).	

		The mechanism to produce the jet is also uncertain.
		We assumed that the jet is driven by the BZ process \citep{BlandfordZnajek1977, McKinney2005}. 
		\cite{MacFadyenWoosley1999, Kawanaka_etal_2013, Liu_etal_2015, Lei_etal_2017}  pointed out that jet luminosity driven by pair annihilation of neutrinos and antineutrinos is less than that of BZ process by a few orders of magnitudes.
		We also estimated the jet energy by the neutrino annihilation instead of BZ process based on the prescription given by \cite{Zalamea_2011}, as coupled with our calculations.
		We then confirmed these previous arguments.
		Indeed, the expected jet energy by the neutrino annihilation may be similar to those derived for low luminosity GRBs.
		As such, this would account for low-luminosity GRBs accompanied with by energetic GRB-SNe.
		This might be possible if the BZ mechanism may work in some situation but not always, e.g., depending on the magnetic field content within the progenitor star.

		In addition, our model does not include possible effects through the jet propagation in the entire envelope.
		When the jet penetrates the envelope, a cocoon would spread laterally and transfer a part of the jet energy to the envelope.
		This would convert the jet energy to the SN energy, while the sum must be kept.
		Our simulation is not able to produce an energetic GRB-SN associated with a low luminosity GRB, such as GRB980425/SN1998bw.
		It has indeed been suggested that these GRB-SNe may be attributed to `failed' GRB, where the jet loses the most of the energy to the envelope (and powering an SN) \citep{Khokhlov_etal_1999,MacFadyen_etal_2001, Huang_etal_2002, Totani_2003, Zhang_etal_2003,Mizuta_etal_2006, Morsony_etal_2007, Bromberg_etal_2011, Lazzati_etal_2013, Lopez_etal_2013, Mizuta_Ioka_2013, Bromberg_etal_2016,Colle_etal_2017}.
		How this process would work once coupled with the present model is another direction for the future work.

\section{Summary and Conclusions}
	
	Since 1998, more than twenty GRB-SNe have been detected \citep{Galamaetal1998,Woosley_Bloom_2006}.
	In this paper, we construct a simple but self-consistent collapsar model, which takes into account the interactions between the BH, the disk, and the envelope.
	We further investigate how the system would lead to various observational counter parts depending on the progenitor mass and angular momentum, if the model sequence is to cover properties of the observed GRB-SNe.

	We found that the ram pressure of the infalling material is indeed a key to understanding the nature of the GRB-SN mechanism.
	A quasi-spherical wind would not drive an energetic GRB-SN. 
	The present study clarified that there are several conditions which must be satisfied by the launch of the wind and subsequence propagation of the envelope, if the collapsar model is to explain the GRB-SNe by the disk wind: (1) the disk wind must be highly collimated at the base, and (2) the accretion must continue even after the launch of the SN explosion.
	This condition is opposite to the `quasi-spherical' geometry of GRB-SNe derived observationally, therefore  places a strong constraint on the mechanism to trigger the GRB-SNe.
	While how these two conditions are simultaneously met will require further investigation, our model are able to explain some global properties and relation inferred for GRB-SNe.

	We predict that various types of observational counterparts can be realized in the context of the collapsar model, if the most rapidly rotating progenitors near the break-up are to explain the energetic GRB-SNe like  GRB030329/SN2003dh and GRB091127/SN2009bz.
	The expected outcome is highly dependent on the angular momentum, while it is insensitive to the progenitor mass.

	The authors thank Akihiro Suzuki, Takashi Nagao, Ryo Sawada, Ryoma Ouchi, and Yudai Suwa for stimulating discussion.
	T.H. acknowledges support by Japan Society for the Promotion of Science (JSPS) as JSPS fellow (DC1).
	The work has been supported by JSPS KAKENHI Grant 17J07512 (T.H.) and 17H02864 (K.M.).

			\bibliographystyle{aasjournal}
			\bibliography{hayakawa}

	\end{document}